\documentclass[%
 aip,
 amsmath,amssymb,
reprint,%
xlinenumbers,
]{revtex4-1}

\usepackage{graphicx}% Include figure files
\usepackage{dcolumn}% Align table columns on decimal point
\usepackage{bm}% bold math
\usepackage[utf8]{inputenc}
\usepackage[T1]{fontenc}
\usepackage{mathptmx}
\usepackage[colorlinks=true,citecolor=blue]{hyperref}
\begin{document}

%\preprint{AIP/123-QED}

\title{Discrete Light Bullets in Passively Mode-Locked Semiconductor Lasers}
% Force line breaks with \\

\author{Thomas G. Seidel} 
\affiliation{Institute for Theoretical Physics, University of M\"unster, Wilhelm-Klemm-Str. 9, D-48149 M\"unster, Germany}

\author{Auro M. Perego}
\affiliation{Aston Institute of Photonic Technologies, Aston University, Birmingham B4 7ET, United Kingdom}

\author{Julien Javaloyes}
\affiliation{Departament de F\'{\i}sica, Universitat de les Illes Balears, \& Institute of Applied Computing and Community Code (IAC-3), C/ Valldemossa
	km 7.5, 07122 Mallorca, Spain}	
\author{Svetlana V. Gurevich}
\email{gurevics@uni-muenster.de}
\affiliation{Institute for Theoretical Physics, University of M\"unster, Wilhelm-Klemm-Str. 9, D-48149 M\"unster, Germany}

\affiliation{Center for Nonlinear Science (CeNoS), University of M\"unster, Corrensstrasse 2, D-48149 M\"unster, Germany}
\affiliation{Departament de F\'{\i}sica, Universitat de les Illes Balears, \& Institute of Applied Computing and Community Code (IAC-3), C/ Valldemossa
	km 7.5, 07122 Mallorca, Spain}


\begin{abstract}
	In this paper, we analyze the formation and dynamical properties of discrete light bullets (dLBs) in an array of passively mode-locked lasers coupled via evanescent fields in a ring geometry. Using a generic model based upon a system of nearest-neighbor coupled Haus master equations we show numerically the existence of dLBs for different coupling strengths. In order to reduce the complexity of the analysis, we approximate the full problem by a reduced set of discrete equations governing the dynamics of the transverse profile of the dLBs. This effective theory allows us to perform a detailed bifurcation analysis via path-continuation methods. In particular, we show the existence of multistable branches of discrete localized states (dLSs), corresponding to different number of active elements in the array. These branches are either independent of each other or are organized into a snaking bifurcation diagram where the width of the dLS grows via a process of successive increase and decrease of the gain. Mechanisms are revealed by which the snaking branches can be created and destroyed as a second parameter, e.g., the linewidth enhancement factor or the coupling strength are varied. For increasing couplings, the existence of moving bright and dark dLSs is also demonstrated. 
	%(whose core is at a higher (lower) intensity than the tails)
\end{abstract}

\maketitle
% \textbf{Light bullets (LBs) are composed of pulses of light that are simultaneously confined in the transverse and propagation directions. These phase invariant localized structures have attracted a lot of interest in the last two decades for both fundamental and practical reasons: They are individually addressable and can be envisioned for three-dimensional optical information storage. In this paper we study the possibility to build \emph{discrete} light bullets (dLBs) in an array of passively mode-locked lasers coupled via evanescent fields. Starting with the generic model based upon a system of nearest-neighbor coupled Haus equations, we derive an effective discrete model describing the dynamics of the transverse profile of the dLBs. We provide a detailed bifurcation analysis of this model and compare the results with direct numerical simulations of the full system.}

\section{Introduction}
Discrete localized states in nonlinear lattices appear in many areas of research such as biological molecular chains or energy transfer in protein $\alpha$-helices~\cite{D-JTB-73,DK-PSSB-73}, conducting polymer chains~\cite{SSH-PRL-79,HKSS-RMP-88}, solid-state systems~\cite{FG-PhysRep-2008}, Bose-Einstein condensate~\cite{TS-PRL-01} or optical wave-guides~\cite{CJ-OL-88,CDLS-Nature-03} just to mention a few. In nonlinear optical systems these states are often referred to as \emph{discrete solitons} (dSs) and they have been a subject of intense investigation in recent years both theoretically and experimentall, see e.g., Ref.~\cite{LSCASS-PhysRep-2008} for a review. In particular, one- and two-dimensional dSs were predicted theoretically and observed experimentally in arrays of weakly coupled nonlinear cavities with Kerr, saturable cubic, and quadratic nonlinearities~\cite{ESMBA-PRL-98,FCSEC-PRL-03,FSENC-Nature-03,ISSPLS-PRL-04,EPL-PRE-05,YCS-PRA-08,YC-SIADS-10,PYJD_OL-12}. 

The properties of dSs usually differ from those of continuous systems. In particular, the lack of translational symmetry in discrete systems usually causes the trapping of dSs by the so-called Peierls-Nabarro potential~\cite{KC-PRE-93} so that they remain at rest unless the coupling between the array elements exceeds some critical value. In the limit of strong coupling, the mobility properties of dSs were discussed in Ref.~\cite{EPL-PRE-05}, whereas in Ref.~\cite{EL-OL-13} the uniformly moving as well as chaotic oscillatory dSs were observed in an array of coupled Kerr-nonlinear cavities. The chimera-like localized states consisting of spatiotemporal chaos embedded in a homogeneous background were recently studied ~\cite{CFCRT-OL-17} in a discrete model for an array of coupled-waveguide resonators subject to optical injection. These intermittent spatiotemporal chaotic states are shown to coexist with stationary dSs corresponding to different numbers of active waveguides. The multistability and snaking behaviour of dSs were also reported in the model for optical cavities with focusing saturable nonlinearity~\cite{YCS-PRA-08,YC-SIADS-10}. The interaction properties of short pulse trains in an array of nearest-neighbor coupled passively mode-locked lasers were recently addressed in Ref.~\cite{PVPGY-PRL-17}. It was shown that this array can produce a periodic train of clusters consisting of two or more closely packed pulses with the possibility to change the interval between them via the variation of the coupling parameter. 

Passive mode locking (PML) is a well-known method for achieving short optical pulses~\cite{haus00rev}. For proper parameters, the combination of a laser amplifier providing gain and a nonlinear loss element, usually a saturable absorber, leads to the emission of temporal pulses much shorter than the cavity round-trip. However, if operated in the so-called long-cavity regime,  the PML pulses become individually addressable temporal localized states coexisting with the off solution~\cite{MJB-PRL-14,CJM-PRA-16,SJG-PRA-18}. In this regime, the round-trip time is much longer than the semiconductor gain recovery time, which is the slowest variable. This temporal confinement regime was found to be compatible with an additional spatial localization mechanism, leading to the formation of stable three-dimensional light bullets, i.e. localized pulses of light that are simultaneously confined in the transverse and propagation directions~\cite{J-PRL-16,GJ-PRA-17,PJG-PTRA-18}. Light bullets have attracted a lot of attention in the last two decades; In particular they should be addressable, i.e., one can envision that they would circulate independently within an optical cavity as elementary bits of information.

\begin{figure}
	\includegraphics[width=0.99\columnwidth]{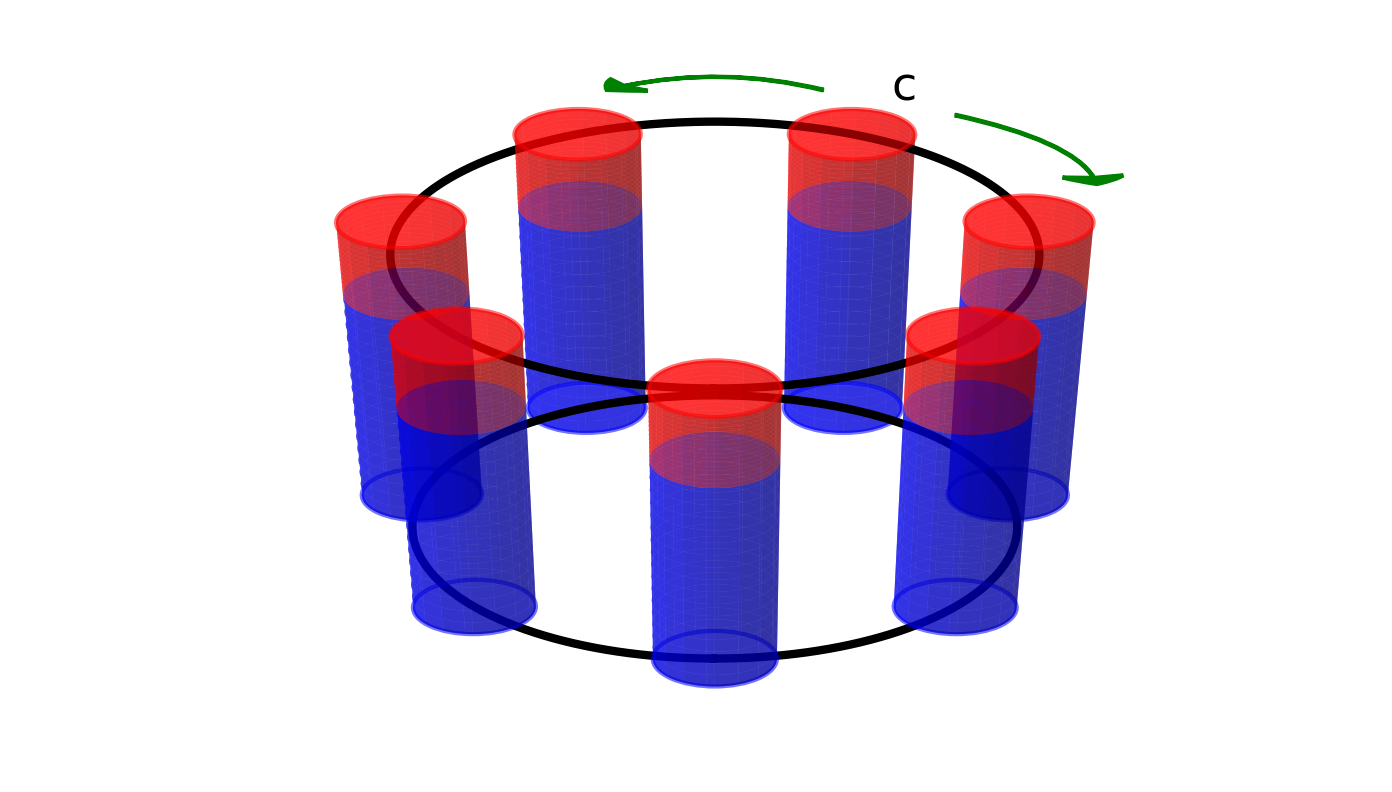}
	\caption{Schematic representation of a ring array of coupled mode-locked lasers. Blue and red parts of each PML laser correspond to the gain and absorber sections, respectively. The arrows indicate the coupling via evanescent fields with the coupling strength $c$. 
}
	\label{fig:model_scetch} 
\end{figure}

In this paper we study the formation and dynamical properties of \emph{discrete light bullets} (dLBs) in an array of PML lasers coupled via evanescent fields in a ring, see Fig.~\ref{fig:model_scetch}.  Here the blue and red parts correspond to the gain and absorber sections of the individual PML laser whereas the arrows indicate the next neighbor coupling with the coupling strength $c$. We perform the analysis in this paper in two steps: First, using an ensemble of nearest-neighbor coupled Haus master equations we show the existence of dLBs for a wide range of coupling strengths. To understand the localization mechanism in details, we approximate the solution of the full system by the product of a slowly evolving discrete transverse profile and of a short temporal pulse propagating inside the cavity. This allows us to obtain a reduced discrete model governing the dynamics of the transverse profile. This effective model termed the \emph{discrete Rosanov equation} allows for a detailed multi-parameter bifurcation study. It also enables us to identify the different mechanisms of instabilities of transverse dLBs. 

\section{Model}

We describe the PML laser array in Fig.~\ref{fig:model_scetch} using nearest-neighbor coupled Haus master equations~\cite{haus00rev,J-PRL-16,GJ-PRA-17} for the evolution of the field profile $E_j=E_j(z,\sigma)$, $j=1,\,\dots\,N$, over the slow time scale $\sigma$ that corresponds to the number of round-trips in the cavity
\begin{eqnarray}
\partial_{\sigma}E_j & = & \left\{ \sqrt{\kappa}\left[1+\frac{1-i\alpha}{2}G_j-\frac{1-i\beta}{2}Q_j\right]-1+\frac{1}{2\gamma^{2}}\partial_{z}^{2}\right\} E_j \nonumber \\
 & + & i\,c\,\left(E_{j-1}+E_{j+1}\right),\label{eq:VTJ1}
\end{eqnarray}
whereas $z$ is a fast time-like variable representing the evolution of the field within the round-trip. The carrier dynamics for $G_j=G_j(z)$ and $Q_j=Q_j(z)$ reads 
\begin{eqnarray}
\partial_{z}G_j & = & \Gamma G_{0}-G_j\left(\Gamma+\left|E_j\right|^{2}\right),\label{eq:VTJ2}\\
\partial_{z}Q_j & = & Q_{0}-Q_j\left(1+s\left|E_j\right|^{2}\right)\,.\label{eq:VTJ3}
\end{eqnarray}
Here, $\kappa$ is the fraction of the power remaining in the cavity after each round-trip, $\gamma$ is the bandwidth of the spectral filter, $\alpha$ and $\beta$ are the linewidth enhancement factors of the gain and absorber sections, respectively and $c$ denotes the nearest neighbor coupling constant. Further, $G_{0}$ is the pumping rate, $\Gamma$ is the gain recovery
rate, $Q_{0}$ is the value of the unsaturated losses, and $s$ the ratio of the saturation energy of the gain and of the saturable absorber sections.

\begin{figure}
	\includegraphics[width=0.99\columnwidth]{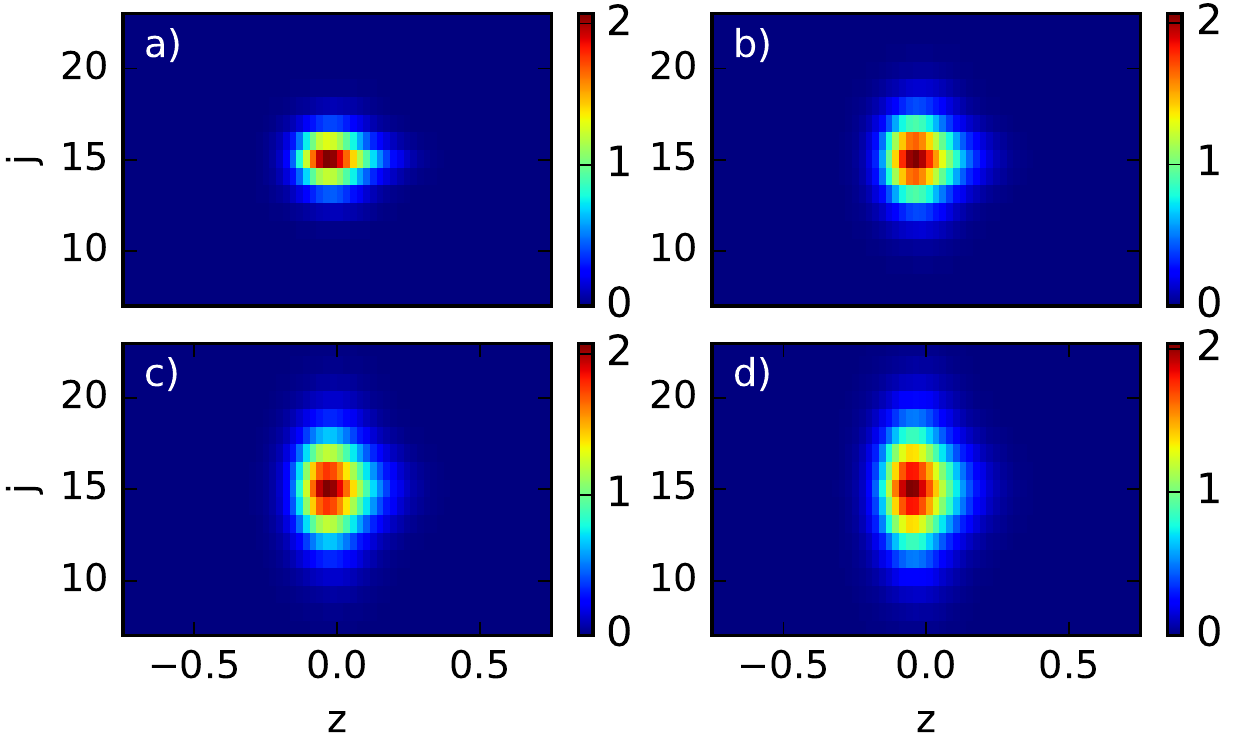}
	\caption{Exemplary solutions of Eqs.~\eqref{eq:VTJ1}-\eqref{eq:VTJ3} showing the intensity profile of a stable dLB existing in an array of $N=30$ elements for different couplings strength $c$: a) c=0.05, b) c=0.10, c) c=0.15, d) c=0.20. Other parameters are: $(\gamma,\kappa,\alpha,\beta,G_0,\Gamma,Q_0,s,L_z,N_z)=(40,0.8,1.5,0.5,0.3840,0.04,0.3,30,5,128)$,where $L_z$ is the length of the cavity and $N_z$ are the number of grid points. If not otherwise stated, all the data represented in the figures are dimensionless.}
	\label{fig:Haus_profiles_dif_c} 
\end{figure}

For proper parameters, Eqs.~\eqref{eq:VTJ1}-\eqref{eq:VTJ3} sustain the existence of stable dLBs as depicted in Fig.~\ref{fig:Haus_profiles_dif_c}. Here, the intensity profile of a stable dLB in the array of $N=30$ elements is shown for four different values of the coupling strength $c$. 

To understand the formation mechanism of dLBs in detail, we start the analysis with the dynamics of the transverse profile of a dLB, that in the following we refer to as \emph{a discrete localized state} (dLS). To this aim we follow~\cite{N-JQE-74,J-PRL-16,GJ-PRA-17} to derive an approximate model governing the shape of the transverse profile. We assume that each field $E_j=E_j(z,\sigma)$ is represented as a product of a short normalized temporal localized pulse $p\left(z\right)$ upon which the dLB is built and  a slowly evolving amplitude of the transverse field $A_j\left(\sigma\right)$, i.e., $E_j(z,\sigma)=p\left(z\right)\,A_j\left(\sigma\right)$. Note that this is a strong approximation as we assume the temporal pulse $p(z)$ to be identical in width and timing for all the array elements. Separating the temporal evolution into the fast and slow parts corresponding to the pulse emission and the subsequent gain recovery allows us to find the discrete equation governing the dynamics of $A_j=A_j(\sigma)$, $j=1,\,\dots\,N$, as 
\begin{equation}
\partial_{\sigma} A_j=i\,c\,(A_{j+1}-2A_j+A_{j-1})+F(|A_j|^2)A_j\,.
\label{eq:Rosa1}
\end{equation}
Defining $h\left(p_j\right)=\left(1-e^{-p_j}\right)/p_j$ with $p_j=|A_j|^{2}$ the nonlinear function $F$ reads
\begin{equation}
F\left(p_j\right)=\sqrt{\kappa}\left[1+\frac{1-i\alpha}{2}G_0\,h(p_j)-\frac{1-i\beta}{2}Q_0\,h(sp_j)\right]\,.\label{eq:Rosa2}
\end{equation}
Note that the continuous counterpart of the discrete equation~\eqref{eq:Rosa1}, obtained taking the limit $c\rightarrow\infty$ with a nonlinear function $F$ corresponding to a static saturated nonlinearity, i.e. $h=1/(1+|A|^2)$ is a so-called Rosanov equation \cite{RK-OS-88,VFK-JOB-99} that is known in the context of static transverse autosolitons in bistable interferometer. 

% We define $G_{th}=\frac{2}{\sqrt{\kappa}}-2+Q_{0}$ as the threshold gain value above which the off solution becomes unstable and introduce  the gain normalized to threshold $g$
% and
% 
% All the localized states are found below the threshold
% for which $G<G_{th}$ or equivalently $g<1$.
%\clearpage

\section{Results}
\begin{figure}
	\includegraphics[width=0.99\columnwidth]{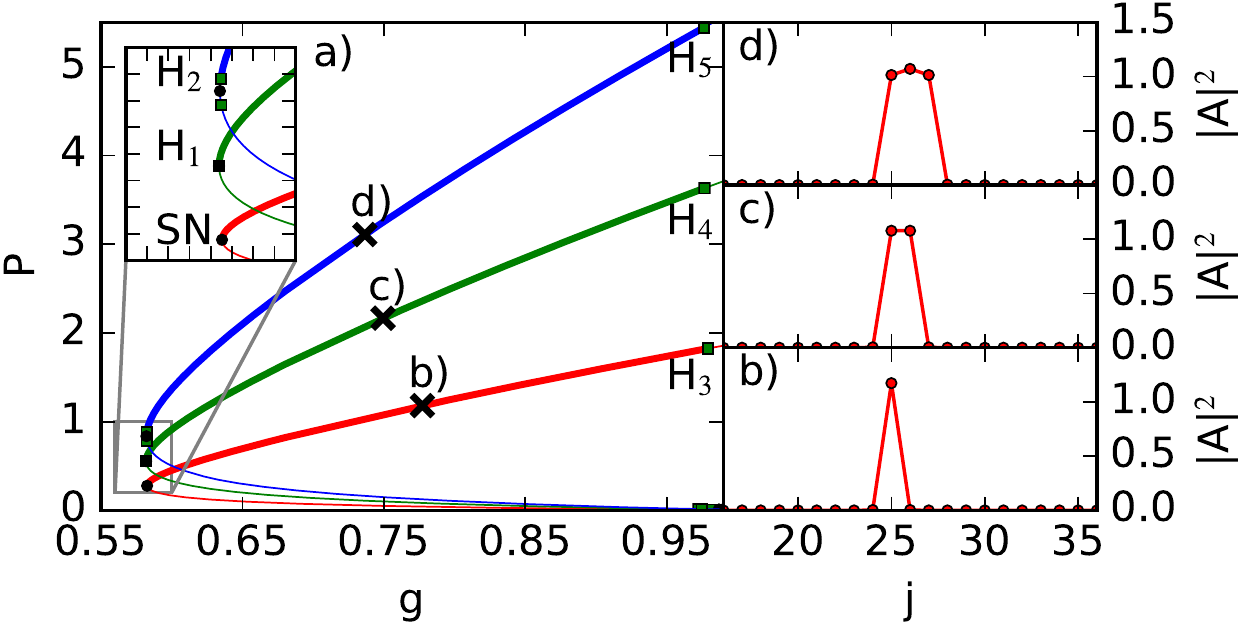}
	\caption{Bifurcation diagram of Eqs.~(\ref{eq:Rosa1},\,\ref{eq:Rosa2}) showing one- (red), two- (green) and three-sites (blue) dLSs. The parameters used are $(Q_0, \alpha, \beta, c, s, \kappa) = (0.3, 1.5, 0.5, 0.004, 30, 0.8)$ and $N=51$ array elements. Thick lines describe stable solution branches, while thin lines stand for unstable ones. Green squares denote Andronov-Hopf (AH) bifurcation points, whereas black dots stand for Saddle-Node (SN) bifurcations. The inset gives a zoomed view on the area around the folds. The panels b), c) and d) show exemplary intensity profiles marked in a).}
	\label{fig:bif_g-123lasing} 
\end{figure}

\begin{figure}
	\includegraphics[width=0.99\columnwidth]{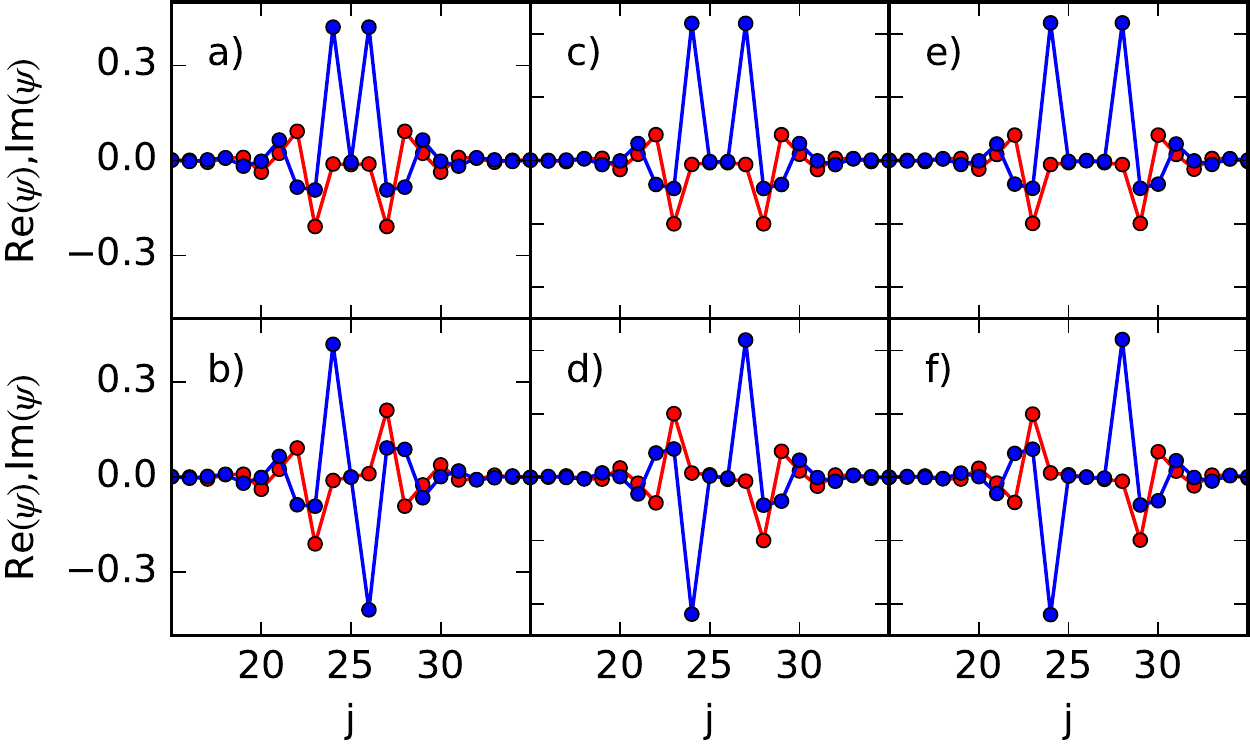}
	\caption{Real and imaginary parts of the critical eigenfunctions $\psi$ of the double Hopf bifurcations H\textsubscript{3} (a,b), H\textsubscript{4} (c,d) and H\textsubscript{5} (e,f) ( cf. Fig. \ref{fig:bif_g-123lasing}). The red points correspond to the $\mathrm{Re}(\psi)$, whereas the blue points to $\mathrm{Im}(\psi)$, respectively.}
	\label{fig:eigenfunctions-Hopf} 
\end{figure}

\begin{figure}[h!]
	\includegraphics[width=0.99\columnwidth]{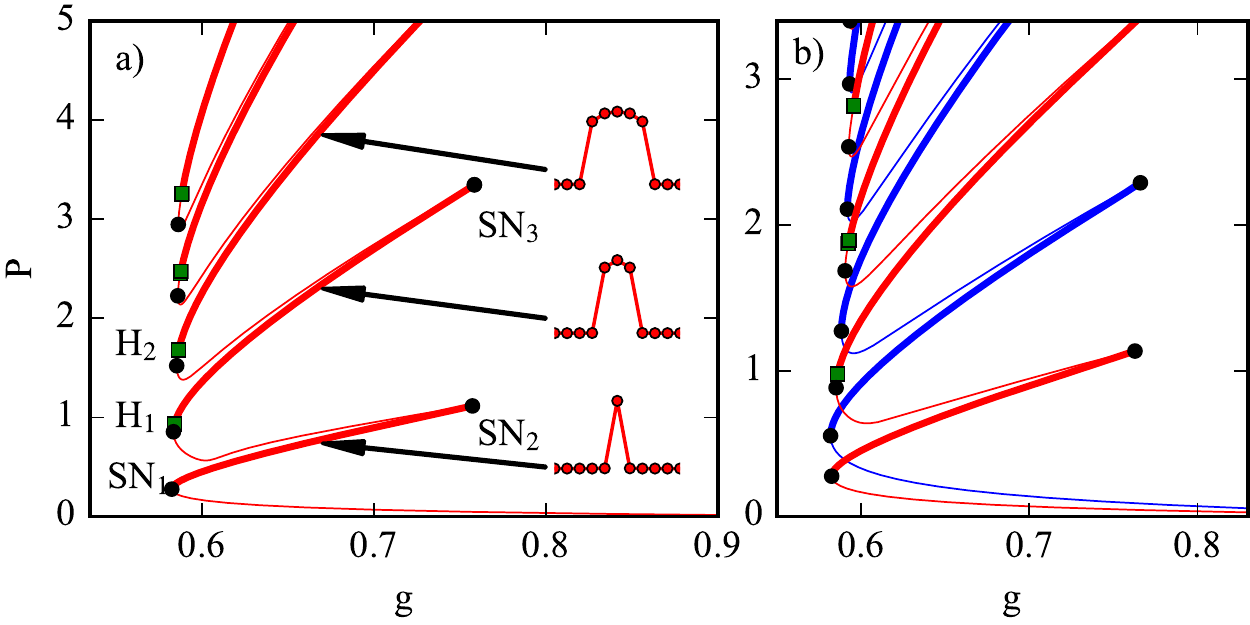}
	\caption{a) Bifurcation diagram in the ($g,P$) plane for $\alpha=0.8$ showing the snaking between the branches of dLSs with different odd number of lasing nodes. The three insets display the solution profiles at $g=0.67$ for the one-, three- and five-site dLSs, respectively. b) Bifurcation diagram for $\alpha=0.6$, where the branches for both odd (red) and even (blue) number of lasing nodes is shown (see the Supplementary Material for the video showing the profile evolution along the branches). Other parameters are: $(Q_0, \beta, c, s, \kappa) = (0.3, 0.5, 0.004, 30, 0.8)$ and N=51.}
	\label{fig:bif_alpha080-snaking} 
\end{figure}

A single solution of Eqs.~(\ref{eq:Rosa1},\,\ref{eq:Rosa2}) can be found in the form 
\begin{equation}
A_j(\sigma)=a_j\,e^{-i\omega \sigma}\,,\label{eq:ansatz}
\end{equation}
where $a_j$ is a complex amplitude of each array element and $\omega$ represents the carrier frequency of
the solution. Substituting Eq.~\eqref{eq:ansatz} into Eqs.~(\ref{eq:Rosa1},\,\ref{eq:Rosa2}) we are left searching for unknowns $a_j$ and $\omega$ of the following algebraic equations set
% The system possesses a phase shift symmetry such that when the power $P=\sum_i |A_i|^2$ reaches a steady state in direct numerical simulations the real and imaginary part oscillate with a certain carrier frequency. In order conduct a continuations in different parameters the symmetry needs to be broken. We use the ansatz
% \begin{equation}
% A_j(t) = a_j e^{-i\omega t},
% \label{eq:ansatz}
% \end{equation}
% where $\omega$ is a so-called frequency shift or a spectral parameter and $a_j$ describes a time independent amplitude. 
\begin{equation}
i\,c\,(a_{j+1}-2a_j+a_{j-1})+i\omega a_j+F(|a_j|^2)a_j=0\,.
\label{eq:differential equations time invariant}
\end{equation}
We followed the solutions of Eqs.~\eqref{eq:differential equations time invariant} in parameter space, by using pseudo-arclength continuation within the AUTO-07P framework~\cite{DKK-I-IJBC-91,DKK-II-IJBC-91,AUTO}. Here, the spectral parameter $\omega$ becomes an additional free parameter that is automatically adapted during the continuation. We define $G_{th}=\frac{2}{\sqrt{\kappa}}-2+Q_{0}$ as the threshold gain value above which the off solution $a_j=0$ for all $j=1,\,\dots\,N$ becomes unstable. The primary continuation parameter could be e.g., the gain normalized to the threshold $g=G_0/G_{th}$, the linewidth enhancement factor $\alpha$ or the coupling strength $c$. 
\paragraph*{Multistability of dLSs} One can start at, e.g., a numerically given solution, continue it in parameter space, and obtain a dLS solution branch. The result for an array of $N=51$ elements is presented in Fig.~\ref{fig:bif_g-123lasing}, where in the panel a) the power $P=\sum_i |a_i|^2$ of three different dLSs is depicted as a function of the normalized gain $g$ for the fixed small coupling strength $c$.
One can see that dLSs only occur in discrete widths corresponding to different numbers of lasing lasers in the array, see Fig.~\ref{fig:bif_g-123lasing}(b,d), where the exemplary profiles of one-, two- and three-sites dLSs are shown. Further, the system is multistable, and we find separate branches for solution profiles containing different number of lasing nodes. Each of the branches bifurcates from the threshold $g=1$, possesses a fold at some fixed value (marked as a black circle in Fig.~\ref{fig:bif_g-123lasing}), and goes to higher intensities. The stability properties of different dLSs branches are similar: The one-site-dLS (red line) is stable between the saddle-node bifurcation point (SN) and the Andronov-Hopf (AH) bifurcation point $H_3$ (marked as a green square) close to the threshold. However, for two- and three-sites dLSs, other AH bifurcations occur around the SN point (e.g., $H_1$, and $H_2$, see inset in Fig.~\ref{fig:bif_g-123lasing}(a) that can limit the stability of the dLSs for low values of the gain. For the higher bias and intensities, the stability is again limited by the AH bifurcations (cf. $H_4$ and $H_5$ points) close to $g=1$ value. Analyzing the destabilizing AH bifurcations on the right flank of the branches reveals that each of $H_3-H_5$ points actually corresponds to a double AH bifurcation. Here, the imaginary parts of the two eigenvalues are the same which means that both eigenmodes exhibit the same frequency. The real and imaginary parts of the corresponding eigenmodes for the $H_3-H_5$ bifurcation points are shown in Fig.~\ref{fig:eigenfunctions-Hopf}. One can see that there is always one even (upper row) and one odd (lower row) eigenmode.
\paragraph*{Snaking bifurcation of dLSs} Interestingly, the bifurcation structure of the branches becomes different if the linewidth enhancement factor $\alpha$ is varied. In particular, reducing $\alpha$ reveals a snaking structure in the bifurcation diagram (see Fig. \ref{fig:bif_alpha080-snaking}). 
Here, the stable parts of the branches for odd, i.e., one-, three-, five-, etc. sites dLS (thick lines) are connected via SN bifurcations by unstable connections (thin lines). The stability on the left side is limited by a SN bifuraction for the one-site (SN\textsubscript{1}, black dot) or AH bifurcations (H\textsubscript{1}, H\textsubscript{2}, green square, for three or more lasing dLSs). For the increasing value of the control parameter $g$, the dLSs become unstable in SN bifurcations (see e.g., SN\textsubscript{2} and SN\textsubscript{3} points). Furthermore, the branches with an odd dLSs are not connected to the ones with an even dLSs, see Fig. \ref{fig:bif_alpha080-snaking}(b), because with increasing $g$ the neighboring nodes of the array are excited symmetrically such that switching from an even number of lasing lasers to an odd number is not possible. Note that the snaking bifurcation of dLSs was also reported in~\cite{YCS-PRA-08, YC-SIADS-10}, where a discrete model for optical cavities with focusing saturable nonlinearity was studied.	

% %
% \begin{figure*}[ht!]
% 	\includegraphics[width=0.99\textwidth]{FIGURES/output/bifurcation_different-alpha.pdf}
% 	\caption{ Bifurcation diagrams in $(g, P)$ plane for different values of $\alpha=(0, 0.15, 0.18, 0.245, 0.25, 0.8, 0.817, 0.9, 1.3)$ for the panels a)-i), respectively. The creation/destruction cycle of the snaking structure is shown. Red and blue branches correspond to one- and three-sites dLS solutions. Thick and thin lines stand for stable and unstable branch parts, respectively. The green branch in e) and f) appears by the reconnection of the one- and three-sites dLS branches and is unstable. See the Supplementary Material section for the visualization of the transition between independent branches and snaking behavior. Other parameters are $(Q_0, \beta, c, s, \kappa, N) = (0.3, 0.5, 0.004, 30, 0.8, 51)$.}
% 	\label{fig:bif_different-alpha} 
% \end{figure*}
% %

\begin{figure*}[ht!]
	\includegraphics[width=0.99\textwidth]{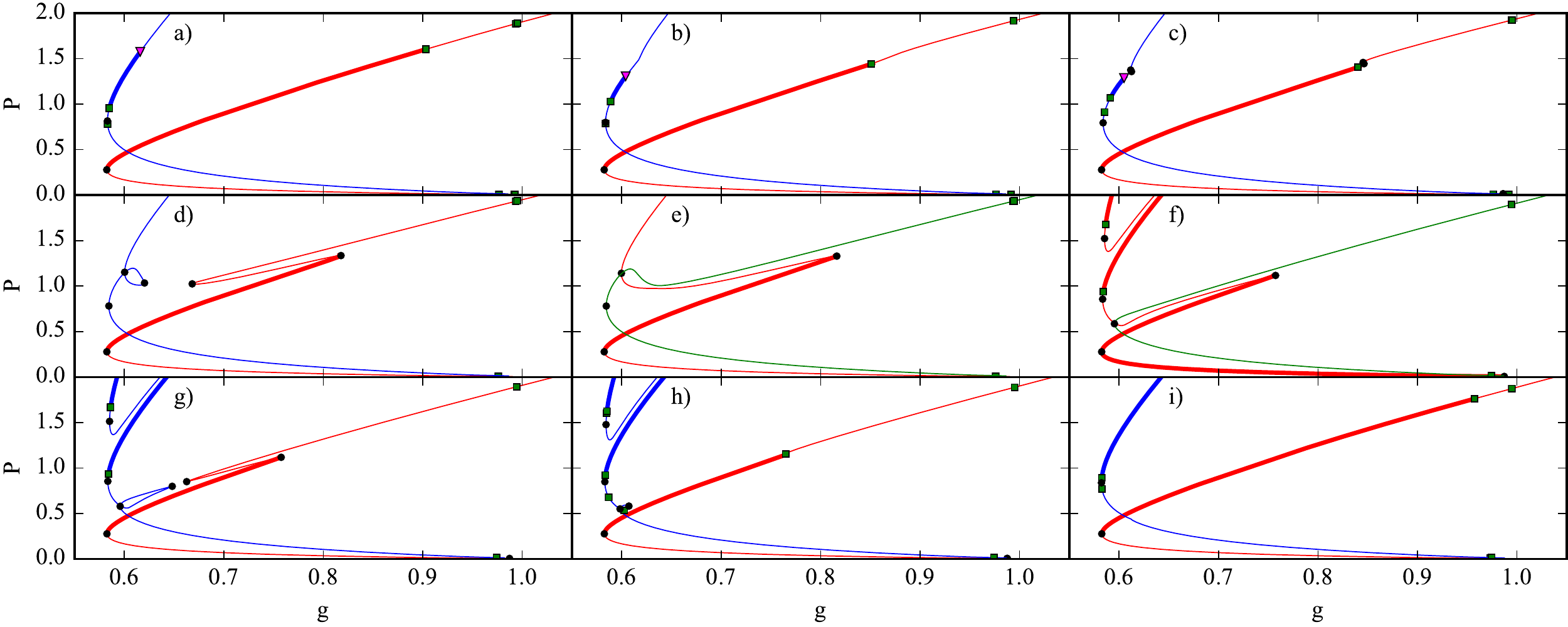}
	\caption{ Bifurcation diagrams in $(g, P)$ plane for different values of $\alpha=(0, 0.15, 0.18, 0.245, 0.25, 0.8, 0.817, 0.9, 1.3)$ for the panels a)-i), respectively. The creation/destruction cycle of the snaking structure is shown. Red and blue branches correspond to one- and three-sites dLS solutions. Thick and thin lines stand for stable and unstable branch parts, respectively. The green branch in e) and f) appears by the reconnection of the one- and three-sites dLS branches and is unstable. See the Supplementary Material section for the visualization of the transition between independent branches and snaking behavior. Other parameters are $(Q_0, \beta, c, s, \kappa, N) = (0.3, 0.5, 0.004, 30, 0.8, 51)$.}
	\label{fig:bif_different-alpha} 
\end{figure*}

\begin{figure}[ht!]
	\includegraphics[width=0.99\columnwidth]{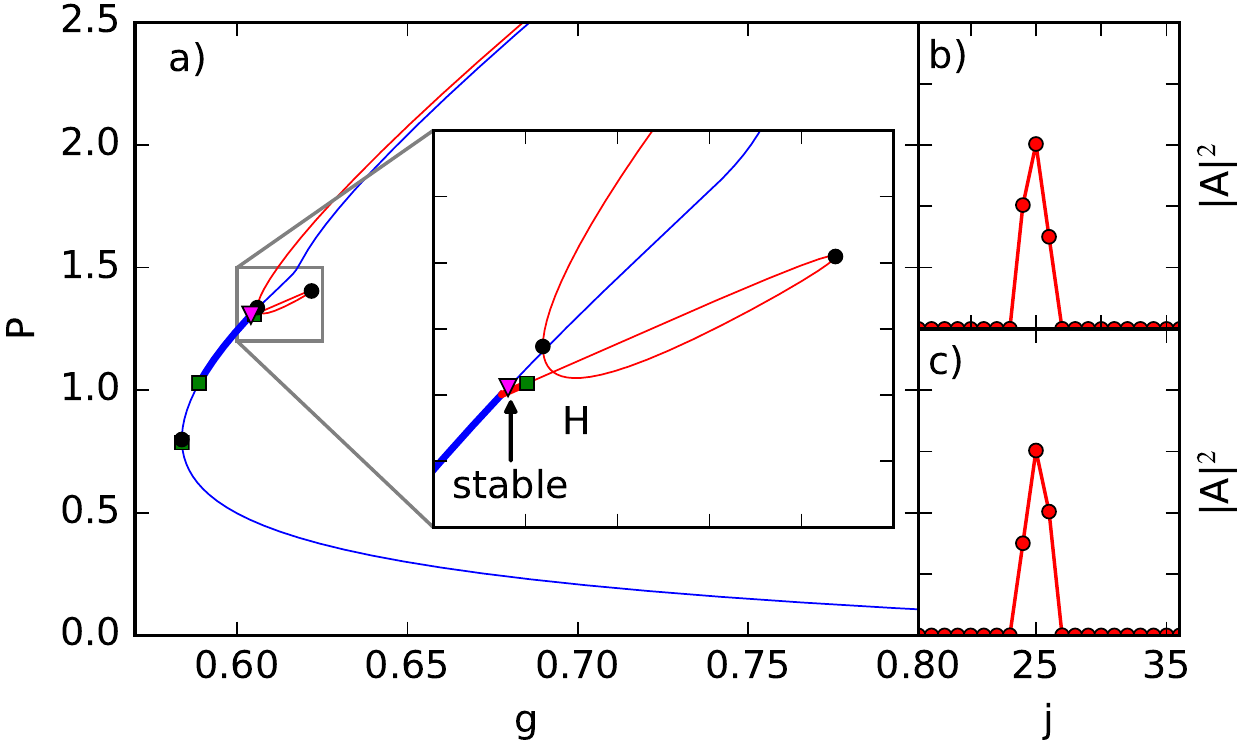}
	\caption{a) Branch of a three-sites dLS in ($g, P$) plane for $\alpha=0.15$ (blue line, cf. Fig.~\ref{fig:bif_different-alpha}~b)). The inset shows a zoom into the area where two branches of unsymmetrical dLSs split off the main branch (red line) in a pitchfork bifurcation (magenta triangle). The asymmetrical left- and right-site dLSs profiles at the AH bifurcation point $H$ are plotted in b) and c). Both branches have the same norm $P$ and coincide in the bifurcation diagram in a) (see the Supplementary Material for a video showing the evolution of the solution profiles along the branch). Other parameters as in Fig.~\ref{fig:bif_different-alpha}.}
	\label{fig:bif_alpha015-unsymmetrical} 
\end{figure}

\paragraph*{Bifurcation analysis of the snaking}
Now we want to understand the transition between the independent branches for different dLSs as presented in Fig.~\ref{fig:bif_g-123lasing} to the snaking structure as shown in Fig.~\ref{fig:bif_alpha080-snaking}. To this aim, we analyze in details the behavior of the branches of dLSs corresponding to different values of $\alpha$. For simplicity here we focus on the transition between the solution profiles with one- and three-sites dLSs and Fig.~\ref{fig:bif_different-alpha} shows the resulting bifurcation diagrams in the $(g, P)$ plane obtained for different $\alpha$. 
Figure~\ref{fig:bif_different-alpha}(a,b) indicates that for small values of $\alpha$ the branches for one-site (red) and three-sites (blue) dLSs are not connected to each other. However, the stability on the branches here is different to those shown in Fig. \ref{fig:bif_g-123lasing}: While a one-site dLS is stable between a SN and an AH bifurcation points, for a three-sites dLS the situation is different. In particular, the three-sites dLS gains the stability in a AH bifurcation after the SN point, and looses the stability in a pitchfork bifurcation (marked as a magenta triangle in Fig.~\ref{fig:bif_different-alpha}(a-c). At the pitchfork bifurcation point two branches (red) corresponding to left- and right-site- \emph{asymmetrical} dLSs emerge (see Fig.~\ref{fig:bif_alpha015-unsymmetrical}). Because both dLSs shown in Fig.~\ref{fig:bif_alpha015-unsymmetrical}(b,c) correspond to the same power $P$, the branches coincide for the norm chosen. One can see that the range of stability for these solutions is very small (thick red line) as the branch looses its stability quickly in an AH bifurcation marked with $H$ in Fig.~\ref{fig:bif_alpha015-unsymmetrical}(a). 
For increasing values of $\alpha$ the bifurcation structure becomes more complicated and two additional SN bifurcations appear on both red and blue branches as shown in Fig.~\ref{fig:bif_different-alpha}(c). These bifurcation pairs separate further from each other with $\alpha$,cf. Fig.~\ref{fig:bif_different-alpha}(d), until two of the SN bifurcations, corresponding to the rightmost fold of the blue, three-sites dLS branch and the leftmost of the red one, corresponding to the one-site dLS solutions, merge. This leads to the reconnection of the lower part of the one-site dLS and the upper part of the three-sites dLS branches and a snaking structure emerges, see the red branch in Fig.~\ref{fig:bif_different-alpha}(e). As this takes place, the residual (upper part of the red branch) connects to the lower part of the blue one and an unstable branch, shown in green in Fig.~\ref{fig:bif_different-alpha}(e) appears. However, at this point the part of the branch corresponding to three-sites dLS is unstable and can be stabilized by an AH bifurcation if $\alpha$ is increased, see Fig.~\ref{fig:bif_different-alpha}(f). Note that further branches corresponding to larger odd number of lasing elements are created in a similar way. However, further increases in $\alpha$ lead to the break-up of the snaking behavior via the same mechanism the branch was created. In particular, for increasing $\alpha$ a part of unstable residual (green) branch goes close to the main snaking branch and hits it at some fixed $\alpha$. This leads to connection of the red and green branches so that two folds appear as shown in Fig.~\ref{fig:bif_different-alpha}(g). Note that during this reconnection, the appearing branch of the one-site dLS (red) becomes separated from the still snaking branch of three- and five-sites dLSs (blue). For even larger $\alpha$ the SN bifurcations annihilate and disappear (Fig.~\ref{fig:bif_different-alpha}(h) and finally separate, independent branches, for one-site and three-sites dLSs are formed Fig.~\ref{fig:bif_different-alpha}(i). Note that the five-sites dLSs branch separates from the three-sites dLSs branch via the same scenario. For more details see the Supplementary Material section where the video showing the creation and destruction of the snaking structure is shown.  

\begin{figure}[ht!]
	\includegraphics[width=0.99\columnwidth]{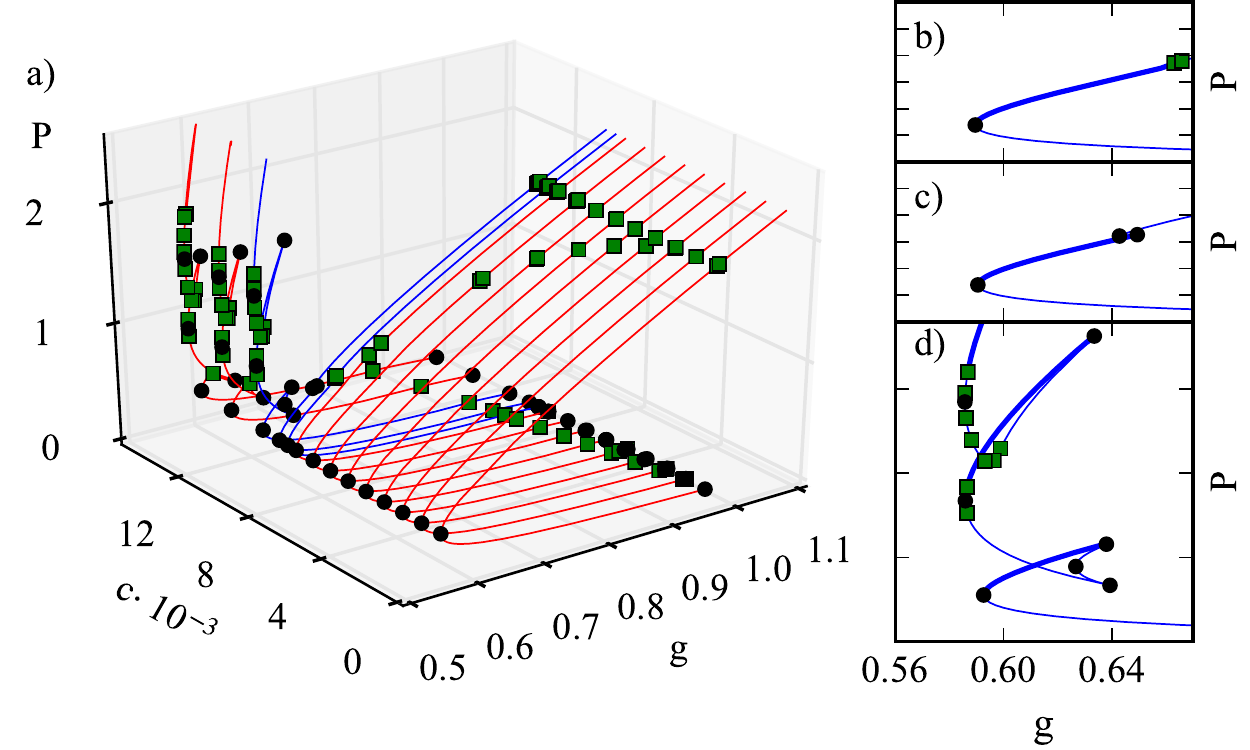}
	\caption{a) Evolution of the branch of the one-site dLS in the $(g, P)$ plane with the coupling strength $c$. The panels b),c) and d) show the branches close to the snaking transition which are marked blue in a) for $c=(0.0095, 0.01, 0.011)$, respectively. Similarly to Fig.~\ref{fig:bif_different-alpha} one can observe snaking branches which occur above the critical coupling $c\simeq 0.011$. In b), c) and d) stable and unstable branches are plotted with thick and thin lines, respectively, whereas black circles and green squares stand for the SN and AH points. Stability information is not displayed in a). Other parameters are $(Q_0, \alpha, \beta, s, \kappa = 0.3, 1.5, 0.5, 30, 0.8)$ and N=51.}
	\label{fig:bif_different-c} 
\end{figure}

\paragraph*{Influence of the coupling strength} Next, we are interested in the influence of the coupling strength $c$ on the dynamics of dLSs. To this aim we reconstructed the branches of the dLSs for different values of $c$ and for the fixed values of $(\alpha, \beta)=(1.5, 0.5)$. In this case the branches of all dLSs are independent, see Fig.~\ref{fig:bif_g-123lasing}. We start with the branch of the one-site dLS and small coupling strength and look to its evolution in $c$. The results are presented in Fig.~\ref{fig:bif_different-c}(a), where the branches for fourteen different values of $c$ are collected in a three-dimensional $(c, g, P)$ bifurcation diagram. 
One can observe that with increasing coupling strength $c$ the branch reveals a similar transition to a snaking as in the case of changing $\alpha$, cf. Fig.~\ref{fig:bif_different-alpha}. In particular, one can see, inspecting Fig.~\ref{fig:bif_different-c}(a) that with increasing $c$ the stability region of the one-site dLS decreases as one of the AH bifurcation points (green square), limiting the stability region is moving towards smaller values of $g$ with increasing $c$. Then, similar to the case of varying $\alpha$, two SN points appear on the branch and the AH point disappears. One of the appearing folds is then connected to the fold of the three-site dLS branch and the snaking bifurcation structure emerges, see Fig.~\ref{fig:bif_different-c}(b-d), where three branches for three values of $c$ illustrating this transition are presented. Here, stable branches for solution profiles corresponding to one-, three-, five-, etc. sites dLSs are interconnected by unstable branches. One can see that with increasing of $c$ new AH bifurcation points appear leading to the decrease of the stability region of small-sites dLSs. This is an expected result as with increasing $c$ the systems tends to the continuous limit where these dLSs do not exist.  

\begin{figure}[ht!]
	\includegraphics[width=0.99\columnwidth]{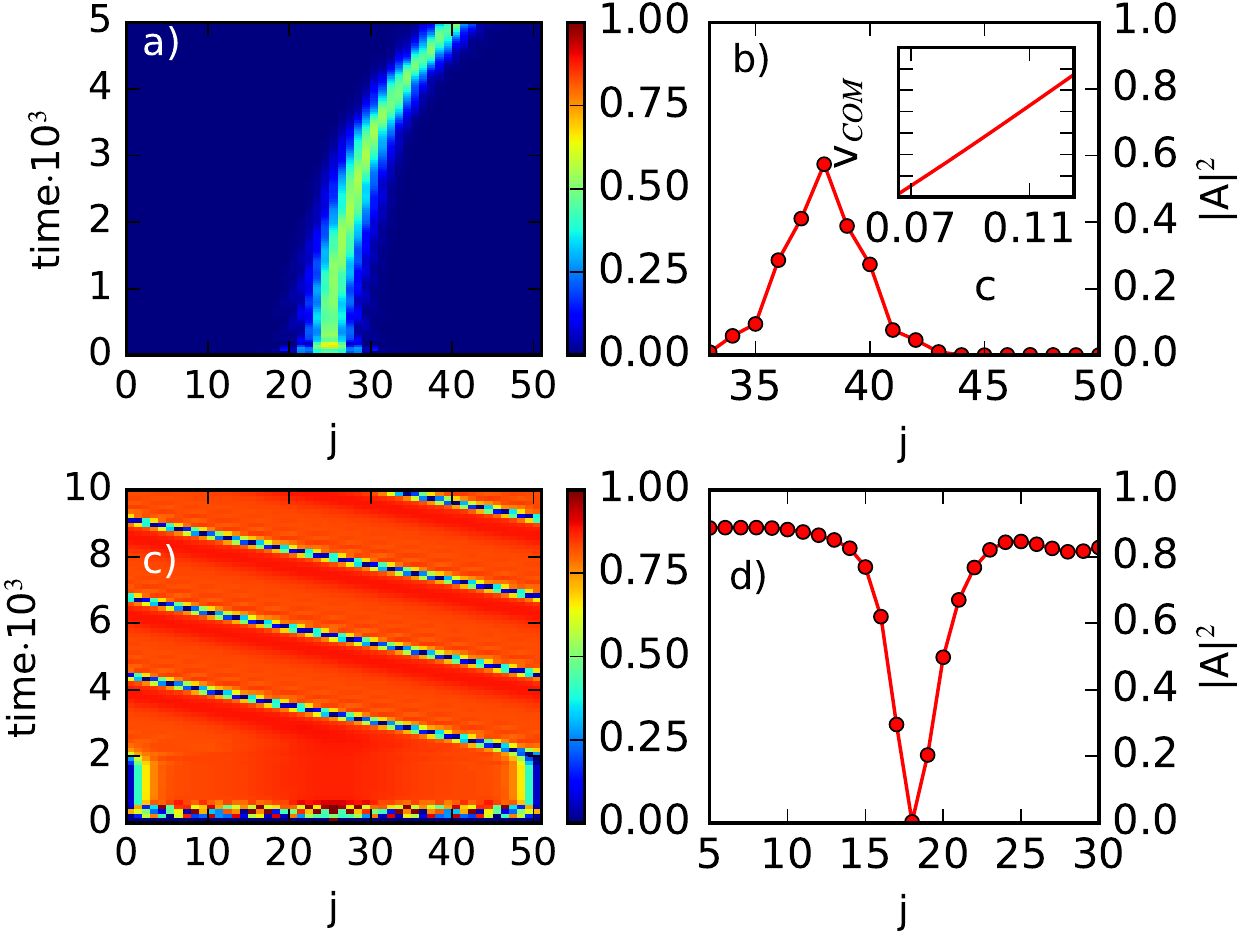}
	\caption{ Time evolution of a bright a) and dark c) moving dLS calculated by direct numerical integration of \eqref{eq:Rosa1},\eqref{eq:Rosa2} for $c=0.1$ and a) $G_0=0.33$ and c)$G_0=0.36$. Panels b) and d) represent the exemplary profiles for both cases at the last time step of the numerical simulation. The inset in b) displays the center of mass velocity $v_{COM}$ as a function of the coupling $c$, showing the linear dependence. See the Supplementary Material for more details of the time evolution. Other parameters are $(Q_0, \alpha, \beta, s, \kappa, N) = (0.3, 1.5, 0.5, 30, 0.8, 51)$.}
	\label{fig:drifting-soliton} 
\end{figure}

\paragraph*{Moving dLSs} 
Finally, we consider the case of even larger coupling strengths. As was mentioned in the Introduction section, the discreteness breaks the translational symmetry that usually causes the trapping of dLSs so that they remain at rest unless e.g., the coupling strength between the nodes exceeds some critical value. An example of a drifting dLS that we refer to as a \emph{bright dLS} in the following, obtained by a direct numerical integration of Eqs.~\eqref{eq:Rosa1}-\eqref{eq:Rosa2} is shown in  Fig.~\ref{fig:drifting-soliton}(a,b). 
A space-time plot is presented in Fig.~\ref{fig:drifting-soliton}(a) where the time evolution of the position of each element $j$ in the array is shown, whereas the color corresponds to the intensity. One can see that after a dLS is formed in the array it becomes unstable, accelerates slowly and drifts to the right. After some time the acceleration phase ends and the dLS moves with a constant velocity. An exemplary solution profile is plotted in Fig.~\ref{fig:drifting-soliton}(b). Note that for the coupling strength used, the dLSs corresponding to smaller number of nodes are unstable. One can demonstrate that the drift velocity of the dLS is determined by the coupling $c$ and the inset in Fig.~\ref{fig:drifting-soliton}(b) clearly shows that the velocity increases linearly with $c$.

Interestingly, besides bright dLSs as shown in Fig.~\ref{fig:drifting-soliton}, the system exhibits also \emph{dark}, or \emph{grey}, moving dLSs, see Fig.~\ref{fig:drifting-soliton}(c,d), which are formed for the same value of $c$ but slightly larger gain values. These dLSs are characterized by one non-lasing node, whereas all other nodes have non-zero intensity. Note that dark moving dLSs were also found in arrays of coupled quadratic nonlinear resonators driven by an inclined holding beam~\cite{EPL-PRE-05} or in coupled in optical cavities with focusing saturable nonlinearity~\cite{YCS-PRA-08}. The time evolution of the dark moving dLS is shown in Fig.~\ref{fig:drifting-soliton}(c). One can see that in the initial phase of the time evolution - because of the strong coupling $c$ and large gain value - more and more laser nodes become unstable until all but one have a non-zero intensity, cf. Fig.~\ref{fig:drifting-soliton}(d). This state remains stationary until the dark dLS spontaneously starts to move with a constant velocity. The motion is facilitated by switching between odd and even number of nodes with zero intensity (see the Supplementary Material section for a video of the moving dark dLS). Furthermore, one sees that the intensity profile is asymmetric and exhibits oscillatory tail on the right side, cf. Fig.~\ref{fig:drifting-soliton}(d). This can potentially lead to the formation of bound states between two dark dLSs in arrays with larger number of elements. Notice that generally the formation mechanisms of the bright and dark moving dLSs are different: The linear stability analysis reveals that for bright dLSs several AH bifurcations trigger the translation while for dark dLS real eigenvalues appear unstable in the spectrum making motion possible.

\paragraph*{Multistability of dLB in the discrete Haus model} 

\begin{figure}[h!]
	\includegraphics[width=0.99\columnwidth]{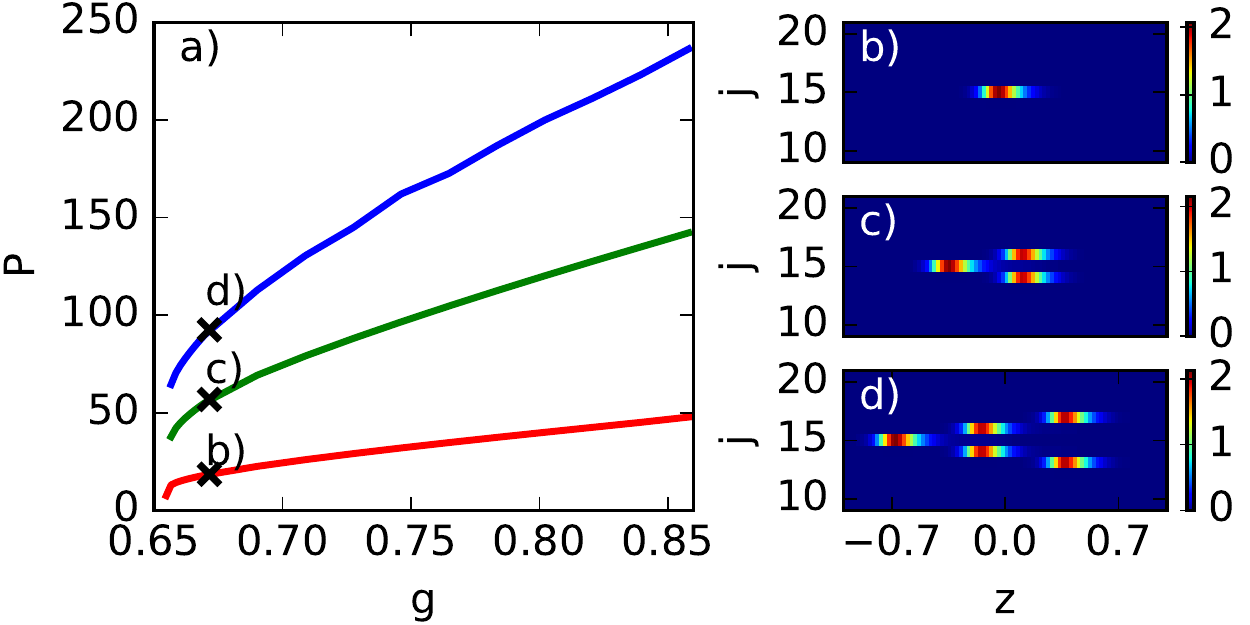}
	\caption{a) Branches of one- (red), three- (green) and five-sites (blue) dLB found by the direct numerical integration of the coupled Haus Eqs.~\eqref{eq:VTJ1}-\eqref{eq:VTJ3} with N=30 array elements. b),c),d): Exemplary profiles of dLBs at g=0.672 (black cross in a)).
	Parameters are $(\gamma,\kappa,\alpha,\beta,T,\Gamma,Q_0,s,c,N_z)=(40,0.8,0.4,0.5,3,0.04,0.3,30,0.004,128)$.}
	\label{fig:bif_Haus} 
\end{figure}

The results of the discrete Rosanov model~\eqref{eq:Rosa1},\eqref{eq:Rosa2} indicate the multistability between different dLSs corresponding to the transverse profile of a dLB. To prove the possible co-existence of different dLBs we go back to the original coupled Haus equations~\eqref{eq:VTJ1}-\eqref{eq:VTJ3} and conduct direct numerical simulations for the case of small coupling $c$, cf. Fig.~\ref{fig:bif_g-123lasing}. We show the resulting branches of one- three- and five-sites dLBs in Fig.~\ref{fig:bif_Haus}.  
One can clearly see that also in the coupled Haus model \eqref{eq:VTJ1}-\eqref{eq:VTJ3} the multistability occurs and dLBs corresponding to different number of odd lasing lasers can form, see Fig. \ref{fig:bif_Haus}(b,d). Note that this scenario also occurs for an even number of nodes.

However, and at variance with the results of Fig.~\ref{fig:Haus_profiles_dif_c}, the small values of the coupling using in Fig.~\ref{fig:bif_Haus} make it so that the individual lasing nodes are separated by a certain offset along the z-axis. This can correspond to the effective repulsive interaction along the fast time axis between individual elements. The latter is induced by the gain dynamics~\cite{KCB-JQE-98,NRV-PD-06,CJM-PRA-16}. Recently it was shown~\cite{PVPGY-PRL-17} that even in the case when the pulses in an individual PML system exhibit strong repulsion, the formation of bound pulse trains can be achieved between the elements of an array of mode-locked lasers coupled via evanescent fields. This way the pulses interact not only within one system but also with those in the neighboring nodes, leading to a different balance between attraction and repulsion. Since the coupled Haus Eqs.~\eqref{eq:VTJ1}-\eqref{eq:VTJ3} can be seen as an effective master equation for the delay differential equation model used in ~\cite{PVPGY-PRL-17} in the long delay limit~\cite{J-PRL-16,CJM-PRA-16,SJG-PRA-18}, the observed dLBs can be interpreted as the fully localized analogues of the periodic train of pulse clusters consisting of two or more closely packed pulses in the array as found in ~\cite{PVPGY-PRL-17}. 

\section{Conclusion}
We studied the formation and the dynamical properties of dLBs in an array of PML lasers coupled via evanescent fields in a ring geometry. Using nearest-neighbor coupled Haus master equations, we demonstrated the existence of dLBs for the wide range of coupling strength. To understand the formation mechanisms in details, the dynamics of dLBs was approximated by a simplified discrete model governing the dynamics of the transverse profile of the dLB, that we called a dLS. This effective discrete Rosanov equation has allowed for a detailed bifurcation analysis. In particular, for small coupling strengths, our results revealed the multistability between branches corresponding to different kind of dLSs with a varying number of active elements. These branches being independent from each other for one parameter set can become connected in the snaking bifurcation structure if one additional parameter, e.g. the linewidth enhancement factor is varied. The reconnection procedure is very involved and several intricate discrete states, including stationary unsymmetrical dLSs were disclosed. 
Furthermore, it was demonstrated that the snaking behavior between different dLSs branches can also be achieved by changing the coupling strength. Moreover, further increasing of the coupling strength was shown to lead to the formation of the moving bright and dark dLSs. Finally, the multistability of several dLBs was demonstrated in the original coupled Haus model. In contrast to the transverse multistable dynamics, where all the temporal pulses were supposed to synchronized for all the array elements, the elements of the resulting dLBs are not in phase because of the repulsive underlying gain dynamics. These dLBs can be seen as a localized version of the periodic train of clusters consisting of closely packed localized pulses reported recently in Ref.~\cite{PVPGY-PRL-17}. There, one could change the interval between individual pulses via the variation of the coupling phase parameter, which is missing in the coupled Haus model~\eqref{eq:VTJ1}-\eqref{eq:VTJ3} as we assumed the coupling to be evanescent. This interesting issue is out of the scope of this paper and will be discussed elsewhere.

\section*{Acknowledgments}
S.G. thanks PRIME programme of the German Academic Exchange Service (DAAD) with funds from the German Federal Ministry of Education and Research (BMBF). J.J. acknowledge the financial support of the MINECO Project MOVELIGHT (PGC2018-099637-B-100 AEI/FEDER UE).


\begin{thebibliography}{36}%
\makeatletter
\providecommand \@ifxundefined [1]{%
 \@ifx{#1\undefined}
}%
\providecommand \@ifnum [1]{%
 \ifnum #1\expandafter \@firstoftwo
 \else \expandafter \@secondoftwo
 \fi
}%
\providecommand \@ifx [1]{%
 \ifx #1\expandafter \@firstoftwo
 \else \expandafter \@secondoftwo
 \fi
}%
\providecommand \natexlab [1]{#1}%
\providecommand \enquote  [1]{``#1''}%
\providecommand \bibnamefont  [1]{#1}%
\providecommand \bibfnamefont [1]{#1}%
\providecommand \citenamefont [1]{#1}%
\providecommand \href@noop [0]{\@secondoftwo}%
\providecommand \href [0]{\begingroup \@sanitize@url \@href}%
\providecommand \@href[1]{\@@startlink{#1}\@@href}%
\providecommand \@@href[1]{\endgroup#1\@@endlink}%
\providecommand \@sanitize@url [0]{\catcode `\\12\catcode `\$12\catcode
  `\&12\catcode `\#12\catcode `\^12\catcode `\_12\catcode `\%12\relax}%
\providecommand \@@startlink[1]{}%
\providecommand \@@endlink[0]{}%
\providecommand \url  [0]{\begingroup\@sanitize@url \@url }%
\providecommand \@url [1]{\endgroup\@href {#1}{\urlprefix }}%
\providecommand \urlprefix  [0]{URL }%
\providecommand \Eprint [0]{\href }%
\providecommand \doibase [0]{http://dx.doi.org/}%
\providecommand \selectlanguage [0]{\@gobble}%
\providecommand \bibinfo  [0]{\@secondoftwo}%
\providecommand \bibfield  [0]{\@secondoftwo}%
\providecommand \translation [1]{[#1]}%
\providecommand \BibitemOpen [0]{}%
\providecommand \bibitemStop [0]{}%
\providecommand \bibitemNoStop [0]{.\EOS\space}%
\providecommand \EOS [0]{\spacefactor3000\relax}%
\providecommand \BibitemShut  [1]{\csname bibitem#1\endcsname}%
\let\auto@bib@innerbib\@empty
%</preamble>
\bibitem [{\citenamefont {Davydov}(1973)}]{D-JTB-73}%
  \BibitemOpen
  \bibfield  {author} {\bibinfo {author} {\bibfnamefont {A.}~\bibnamefont
  {Davydov}},\ }\href {\doibase https://doi.org/10.1016/0022-5193(73)90256-7}
  {\bibfield  {journal} {\bibinfo  {journal} {Journal of Theoretical Biology}\
  }\textbf {\bibinfo {volume} {38}},\ \bibinfo {pages} {559 } (\bibinfo {year}
  {1973})}\BibitemShut {NoStop}%
\bibitem [{\citenamefont {Davydov}\ and\ \citenamefont
  {Kislukha}(1973)}]{DK-PSSB-73}%
  \BibitemOpen
  \bibfield  {author} {\bibinfo {author} {\bibfnamefont {A.~S.}\ \bibnamefont
  {Davydov}}\ and\ \bibinfo {author} {\bibfnamefont {N.~I.}\ \bibnamefont
  {Kislukha}},\ }\href {\doibase 10.1002/pssb.2220590212} {\bibfield  {journal}
  {\bibinfo  {journal} {physica status solidi (b)}\ }\textbf {\bibinfo {volume}
  {59}},\ \bibinfo {pages} {465} (\bibinfo {year} {1973})},\ \Eprint
  {http://arxiv.org/abs/https://onlinelibrary.wiley.com/doi/pdf/10.1002/pssb.2220590212}
  {https://onlinelibrary.wiley.com/doi/pdf/10.1002/pssb.2220590212}
  \BibitemShut {NoStop}%
\bibitem [{\citenamefont {Su}\ \emph {et~al.}(1979)\citenamefont {Su},
  \citenamefont {Schrieffer},\ and\ \citenamefont {Heeger}}]{SSH-PRL-79}%
  \BibitemOpen
  \bibfield  {author} {\bibinfo {author} {\bibfnamefont {W.~P.}\ \bibnamefont
  {Su}}, \bibinfo {author} {\bibfnamefont {J.~R.}\ \bibnamefont {Schrieffer}},
  \ and\ \bibinfo {author} {\bibfnamefont {A.~J.}\ \bibnamefont {Heeger}},\
  }\href {\doibase 10.1103/PhysRevLett.42.1698} {\bibfield  {journal} {\bibinfo
   {journal} {Phys. Rev. Lett.}\ }\textbf {\bibinfo {volume} {42}},\ \bibinfo
  {pages} {1698} (\bibinfo {year} {1979})}\BibitemShut {NoStop}%
\bibitem [{\citenamefont {Heeger}\ \emph {et~al.}(1988)\citenamefont {Heeger},
  \citenamefont {Kivelson}, \citenamefont {Schrieffer},\ and\ \citenamefont
  {Su}}]{HKSS-RMP-88}%
  \BibitemOpen
  \bibfield  {author} {\bibinfo {author} {\bibfnamefont {A.~J.}\ \bibnamefont
  {Heeger}}, \bibinfo {author} {\bibfnamefont {S.}~\bibnamefont {Kivelson}},
  \bibinfo {author} {\bibfnamefont {J.~R.}\ \bibnamefont {Schrieffer}}, \ and\
  \bibinfo {author} {\bibfnamefont {W.~P.}\ \bibnamefont {Su}},\ }\href
  {\doibase 10.1103/RevModPhys.60.781} {\bibfield  {journal} {\bibinfo
  {journal} {Rev. Mod. Phys.}\ }\textbf {\bibinfo {volume} {60}},\ \bibinfo
  {pages} {781} (\bibinfo {year} {1988})}\BibitemShut {NoStop}%
\bibitem [{\citenamefont {Flach}\ and\ \citenamefont
  {Gorbach}(2008)}]{FG-PhysRep-2008}%
  \BibitemOpen
  \bibfield  {author} {\bibinfo {author} {\bibfnamefont {S.}~\bibnamefont
  {Flach}}\ and\ \bibinfo {author} {\bibfnamefont {A.~V.}\ \bibnamefont
  {Gorbach}},\ }\href {\doibase https://doi.org/10.1016/j.physrep.2008.05.002}
  {\bibfield  {journal} {\bibinfo  {journal} {Physics Reports}\ }\textbf
  {\bibinfo {volume} {467}},\ \bibinfo {pages} {1 } (\bibinfo {year}
  {2008})}\BibitemShut {NoStop}%
\bibitem [{\citenamefont {Trombettoni}\ and\ \citenamefont
  {Smerzi}(2001)}]{TS-PRL-01}%
  \BibitemOpen
  \bibfield  {author} {\bibinfo {author} {\bibfnamefont {A.}~\bibnamefont
  {Trombettoni}}\ and\ \bibinfo {author} {\bibfnamefont {A.}~\bibnamefont
  {Smerzi}},\ }\href {\doibase 10.1103/PhysRevLett.86.2353} {\bibfield
  {journal} {\bibinfo  {journal} {Phys. Rev. Lett.}\ }\textbf {\bibinfo
  {volume} {86}},\ \bibinfo {pages} {2353} (\bibinfo {year}
  {2001})}\BibitemShut {NoStop}%
\bibitem [{\citenamefont {Christodoulides}\ and\ \citenamefont
  {Joseph}(1988)}]{CJ-OL-88}%
  \BibitemOpen
  \bibfield  {author} {\bibinfo {author} {\bibfnamefont {D.~N.}\ \bibnamefont
  {Christodoulides}}\ and\ \bibinfo {author} {\bibfnamefont {R.~I.}\
  \bibnamefont {Joseph}},\ }\href {\doibase 10.1364/OL.13.000794} {\bibfield
  {journal} {\bibinfo  {journal} {Opt. Lett.}\ }\textbf {\bibinfo {volume}
  {13}},\ \bibinfo {pages} {794} (\bibinfo {year} {1988})}\BibitemShut
  {NoStop}%
\bibitem [{\citenamefont {Christodoulides}\ \emph {et~al.}(2003)\citenamefont
  {Christodoulides}, \citenamefont {Lederer},\ and\ \citenamefont
  {Silberberg}}]{CDLS-Nature-03}%
  \BibitemOpen
  \bibfield  {author} {\bibinfo {author} {\bibfnamefont {D.~N.}\ \bibnamefont
  {Christodoulides}}, \bibinfo {author} {\bibfnamefont {F.}~\bibnamefont
  {Lederer}}, \ and\ \bibinfo {author} {\bibfnamefont {Y.}~\bibnamefont
  {Silberberg}},\ }\href@noop {} {\bibfield  {journal} {\bibinfo  {journal}
  {Nature}\ }\textbf {\bibinfo {volume} {424}},\ \bibinfo {pages} {817}
  (\bibinfo {year} {2003})}\BibitemShut {NoStop}%
\bibitem [{\citenamefont {Lederer}\ \emph {et~al.}(2008)\citenamefont
  {Lederer}, \citenamefont {Stegeman}, \citenamefont {Christodoulides},
  \citenamefont {Assanto}, \citenamefont {Segev},\ and\ \citenamefont
  {Silberberg}}]{LSCASS-PhysRep-2008}%
  \BibitemOpen
  \bibfield  {author} {\bibinfo {author} {\bibfnamefont {F.}~\bibnamefont
  {Lederer}}, \bibinfo {author} {\bibfnamefont {G.~I.}\ \bibnamefont
  {Stegeman}}, \bibinfo {author} {\bibfnamefont {D.~N.}\ \bibnamefont
  {Christodoulides}}, \bibinfo {author} {\bibfnamefont {G.}~\bibnamefont
  {Assanto}}, \bibinfo {author} {\bibfnamefont {M.}~\bibnamefont {Segev}}, \
  and\ \bibinfo {author} {\bibfnamefont {Y.}~\bibnamefont {Silberberg}},\
  }\href {\doibase https://doi.org/10.1016/j.physrep.2008.04.004} {\bibfield
  {journal} {\bibinfo  {journal} {Physics Reports}\ }\textbf {\bibinfo {volume}
  {463}},\ \bibinfo {pages} {1 } (\bibinfo {year} {2008})}\BibitemShut
  {NoStop}%
\bibitem [{\citenamefont {Eisenberg}\ \emph {et~al.}(1998)\citenamefont
  {Eisenberg}, \citenamefont {Silberberg}, \citenamefont {Morandotti},
  \citenamefont {Boyd},\ and\ \citenamefont {Aitchison}}]{ESMBA-PRL-98}%
  \BibitemOpen
  \bibfield  {author} {\bibinfo {author} {\bibfnamefont {H.~S.}\ \bibnamefont
  {Eisenberg}}, \bibinfo {author} {\bibfnamefont {Y.}~\bibnamefont
  {Silberberg}}, \bibinfo {author} {\bibfnamefont {R.}~\bibnamefont
  {Morandotti}}, \bibinfo {author} {\bibfnamefont {A.~R.}\ \bibnamefont
  {Boyd}}, \ and\ \bibinfo {author} {\bibfnamefont {J.~S.}\ \bibnamefont
  {Aitchison}},\ }\href {\doibase 10.1103/PhysRevLett.81.3383} {\bibfield
  {journal} {\bibinfo  {journal} {Phys. Rev. Lett.}\ }\textbf {\bibinfo
  {volume} {81}},\ \bibinfo {pages} {3383} (\bibinfo {year}
  {1998})}\BibitemShut {NoStop}%
\bibitem [{\citenamefont {Fleischer}\ \emph
  {et~al.}(2003{\natexlab{a}})\citenamefont {Fleischer}, \citenamefont
  {Carmon}, \citenamefont {Segev}, \citenamefont {Efremidis},\ and\
  \citenamefont {Christodoulides}}]{FCSEC-PRL-03}%
  \BibitemOpen
  \bibfield  {author} {\bibinfo {author} {\bibfnamefont {J.~W.}\ \bibnamefont
  {Fleischer}}, \bibinfo {author} {\bibfnamefont {T.}~\bibnamefont {Carmon}},
  \bibinfo {author} {\bibfnamefont {M.}~\bibnamefont {Segev}}, \bibinfo
  {author} {\bibfnamefont {N.~K.}\ \bibnamefont {Efremidis}}, \ and\ \bibinfo
  {author} {\bibfnamefont {D.~N.}\ \bibnamefont {Christodoulides}},\ }\href
  {\doibase 10.1103/PhysRevLett.90.023902} {\bibfield  {journal} {\bibinfo
  {journal} {Phys. Rev. Lett.}\ }\textbf {\bibinfo {volume} {90}},\ \bibinfo
  {pages} {023902} (\bibinfo {year} {2003}{\natexlab{a}})}\BibitemShut
  {NoStop}%
\bibitem [{\citenamefont {Fleischer}\ \emph
  {et~al.}(2003{\natexlab{b}})\citenamefont {Fleischer}, \citenamefont {Segev},
  \citenamefont {Efremidis},\ and\ \citenamefont
  {Christodoulides}}]{FSENC-Nature-03}%
  \BibitemOpen
  \bibfield  {author} {\bibinfo {author} {\bibfnamefont {J.~W.}\ \bibnamefont
  {Fleischer}}, \bibinfo {author} {\bibfnamefont {M.}~\bibnamefont {Segev}},
  \bibinfo {author} {\bibfnamefont {N.~K.}\ \bibnamefont {Efremidis}}, \ and\
  \bibinfo {author} {\bibfnamefont {D.~N.}\ \bibnamefont {Christodoulides}},\
  }\href@noop {} {\bibfield  {journal} {\bibinfo  {journal} {Nature}\ }\textbf
  {\bibinfo {volume} {422}},\ \bibinfo {pages} {147} (\bibinfo {year}
  {2003}{\natexlab{b}})}\BibitemShut {NoStop}%
\bibitem [{\citenamefont {Iwanow}\ \emph {et~al.}(2004)\citenamefont {Iwanow},
  \citenamefont {Schiek}, \citenamefont {Stegeman}, \citenamefont {Pertsch},
  \citenamefont {Lederer}, \citenamefont {Min},\ and\ \citenamefont
  {Sohler}}]{ISSPLS-PRL-04}%
  \BibitemOpen
  \bibfield  {author} {\bibinfo {author} {\bibfnamefont {R.}~\bibnamefont
  {Iwanow}}, \bibinfo {author} {\bibfnamefont {R.}~\bibnamefont {Schiek}},
  \bibinfo {author} {\bibfnamefont {G.~I.}\ \bibnamefont {Stegeman}}, \bibinfo
  {author} {\bibfnamefont {T.}~\bibnamefont {Pertsch}}, \bibinfo {author}
  {\bibfnamefont {F.}~\bibnamefont {Lederer}}, \bibinfo {author} {\bibfnamefont
  {Y.}~\bibnamefont {Min}}, \ and\ \bibinfo {author} {\bibfnamefont
  {W.}~\bibnamefont {Sohler}},\ }\href {\doibase 10.1103/PhysRevLett.93.113902}
  {\bibfield  {journal} {\bibinfo  {journal} {Phys. Rev. Lett.}\ }\textbf
  {\bibinfo {volume} {93}},\ \bibinfo {pages} {113902} (\bibinfo {year}
  {2004})}\BibitemShut {NoStop}%
\bibitem [{\citenamefont {Egorov}\ \emph {et~al.}(2005)\citenamefont {Egorov},
  \citenamefont {Peschel},\ and\ \citenamefont {Lederer}}]{EPL-PRE-05}%
  \BibitemOpen
  \bibfield  {author} {\bibinfo {author} {\bibfnamefont {O.}~\bibnamefont
  {Egorov}}, \bibinfo {author} {\bibfnamefont {U.}~\bibnamefont {Peschel}}, \
  and\ \bibinfo {author} {\bibfnamefont {F.}~\bibnamefont {Lederer}},\ }\href
  {\doibase 10.1103/PhysRevE.72.066603} {\bibfield  {journal} {\bibinfo
  {journal} {Phys. Rev. E}\ }\textbf {\bibinfo {volume} {72}},\ \bibinfo
  {pages} {066603} (\bibinfo {year} {2005})}\BibitemShut {NoStop}%
\bibitem [{\citenamefont {Yulin}\ \emph {et~al.}(2008)\citenamefont {Yulin},
  \citenamefont {Champneys},\ and\ \citenamefont {Skryabin}}]{YCS-PRA-08}%
  \BibitemOpen
  \bibfield  {author} {\bibinfo {author} {\bibfnamefont {A.~V.}\ \bibnamefont
  {Yulin}}, \bibinfo {author} {\bibfnamefont {A.~R.}\ \bibnamefont
  {Champneys}}, \ and\ \bibinfo {author} {\bibfnamefont {D.~V.}\ \bibnamefont
  {Skryabin}},\ }\href {\doibase 10.1103/PhysRevA.78.011804} {\bibfield
  {journal} {\bibinfo  {journal} {Phys. Rev. A}\ }\textbf {\bibinfo {volume}
  {78}},\ \bibinfo {pages} {011804} (\bibinfo {year} {2008})}\BibitemShut
  {NoStop}%
\bibitem [{\citenamefont {Yulin}\ and\ \citenamefont
  {Champneys}(2010)}]{YC-SIADS-10}%
  \BibitemOpen
  \bibfield  {author} {\bibinfo {author} {\bibfnamefont {A.~V.}\ \bibnamefont
  {Yulin}}\ and\ \bibinfo {author} {\bibfnamefont {A.~R.}\ \bibnamefont
  {Champneys}},\ }\href {\doibase 10.1137/080734297} {\bibfield  {journal}
  {\bibinfo  {journal} {SIAM Journal on Applied Dynamical Systems}\ }\textbf
  {\bibinfo {volume} {9}},\ \bibinfo {pages} {391} (\bibinfo {year} {2010})},\
  \Eprint {http://arxiv.org/abs/https://doi.org/10.1137/080734297}
  {https://doi.org/10.1137/080734297} \BibitemShut {NoStop}%
\bibitem [{\citenamefont {Prilepsky}\ \emph {et~al.}(2012)\citenamefont
  {Prilepsky}, \citenamefont {Yulin}, \citenamefont {Johansson},\ and\
  \citenamefont {Derevyanko}}]{PYJD_OL-12}%
  \BibitemOpen
  \bibfield  {author} {\bibinfo {author} {\bibfnamefont {J.~E.}\ \bibnamefont
  {Prilepsky}}, \bibinfo {author} {\bibfnamefont {A.~V.}\ \bibnamefont
  {Yulin}}, \bibinfo {author} {\bibfnamefont {M.}~\bibnamefont {Johansson}}, \
  and\ \bibinfo {author} {\bibfnamefont {S.~A.}\ \bibnamefont {Derevyanko}},\
  }\href {\doibase 10.1364/OL.37.004600} {\bibfield  {journal} {\bibinfo
  {journal} {Opt. Lett.}\ }\textbf {\bibinfo {volume} {37}},\ \bibinfo {pages}
  {4600} (\bibinfo {year} {2012})}\BibitemShut {NoStop}%
\bibitem [{\citenamefont {Kivshar}\ and\ \citenamefont
  {Campbell}(1993)}]{KC-PRE-93}%
  \BibitemOpen
  \bibfield  {author} {\bibinfo {author} {\bibfnamefont {Y.~S.}\ \bibnamefont
  {Kivshar}}\ and\ \bibinfo {author} {\bibfnamefont {D.~K.}\ \bibnamefont
  {Campbell}},\ }\href {\doibase 10.1103/PhysRevE.48.3077} {\bibfield
  {journal} {\bibinfo  {journal} {Phys. Rev. E}\ }\textbf {\bibinfo {volume}
  {48}},\ \bibinfo {pages} {3077} (\bibinfo {year} {1993})}\BibitemShut
  {NoStop}%
\bibitem [{\citenamefont {Egorov}\ and\ \citenamefont
  {Lederer}(2013)}]{EL-OL-13}%
  \BibitemOpen
  \bibfield  {author} {\bibinfo {author} {\bibfnamefont {O.~A.}\ \bibnamefont
  {Egorov}}\ and\ \bibinfo {author} {\bibfnamefont {F.}~\bibnamefont
  {Lederer}},\ }\href {\doibase 10.1364/OL.38.001010} {\bibfield  {journal}
  {\bibinfo  {journal} {Opt. Lett.}\ }\textbf {\bibinfo {volume} {38}},\
  \bibinfo {pages} {1010} (\bibinfo {year} {2013})}\BibitemShut {NoStop}%
\bibitem [{\citenamefont {Clerc}\ \emph {et~al.}(2017)\citenamefont {Clerc},
  \citenamefont {Ferr\'{e}}, \citenamefont {Coulibaly}, \citenamefont {Rojas},\
  and\ \citenamefont {Tlidi}}]{CFCRT-OL-17}%
  \BibitemOpen
  \bibfield  {author} {\bibinfo {author} {\bibfnamefont {M.~G.}\ \bibnamefont
  {Clerc}}, \bibinfo {author} {\bibfnamefont {M.~A.}\ \bibnamefont
  {Ferr\'{e}}}, \bibinfo {author} {\bibfnamefont {S.}~\bibnamefont
  {Coulibaly}}, \bibinfo {author} {\bibfnamefont {R.~G.}\ \bibnamefont
  {Rojas}}, \ and\ \bibinfo {author} {\bibfnamefont {M.}~\bibnamefont
  {Tlidi}},\ }\href {\doibase 10.1364/OL.42.002906} {\bibfield  {journal}
  {\bibinfo  {journal} {Opt. Lett.}\ }\textbf {\bibinfo {volume} {42}},\
  \bibinfo {pages} {2906} (\bibinfo {year} {2017})}\BibitemShut {NoStop}%
\bibitem [{\citenamefont {Puzyrev}\ \emph {et~al.}(2017)\citenamefont
  {Puzyrev}, \citenamefont {Vladimirov}, \citenamefont {Pimenov}, \citenamefont
  {Gurevich},\ and\ \citenamefont {Yanchuk}}]{PVPGY-PRL-17}%
  \BibitemOpen
  \bibfield  {author} {\bibinfo {author} {\bibfnamefont {D.}~\bibnamefont
  {Puzyrev}}, \bibinfo {author} {\bibfnamefont {A.~G.}\ \bibnamefont
  {Vladimirov}}, \bibinfo {author} {\bibfnamefont {A.}~\bibnamefont {Pimenov}},
  \bibinfo {author} {\bibfnamefont {S.~V.}\ \bibnamefont {Gurevich}}, \ and\
  \bibinfo {author} {\bibfnamefont {S.}~\bibnamefont {Yanchuk}},\ }\href
  {\doibase 10.1103/PhysRevLett.119.163901} {\bibfield  {journal} {\bibinfo
  {journal} {Phys. Rev. Lett.}\ }\textbf {\bibinfo {volume} {119}},\ \bibinfo
  {pages} {163901} (\bibinfo {year} {2017})}\BibitemShut {NoStop}%
\bibitem [{\citenamefont {Haus}(2000)}]{haus00rev}%
  \BibitemOpen
  \bibfield  {author} {\bibinfo {author} {\bibfnamefont {H.~A.}\ \bibnamefont
  {Haus}},\ }\href@noop {} {\bibfield  {journal} {\bibinfo  {journal} {IEEE J.
  Selected Topics Quantum Electron.}\ }\textbf {\bibinfo {volume} {6}},\
  \bibinfo {pages} {1173} (\bibinfo {year} {2000})}\BibitemShut {NoStop}%
\bibitem [{\citenamefont {Marconi}\ \emph {et~al.}(2014)\citenamefont
  {Marconi}, \citenamefont {Javaloyes}, \citenamefont {Balle},\ and\
  \citenamefont {Giudici}}]{MJB-PRL-14}%
  \BibitemOpen
  \bibfield  {author} {\bibinfo {author} {\bibfnamefont {M.}~\bibnamefont
  {Marconi}}, \bibinfo {author} {\bibfnamefont {J.}~\bibnamefont {Javaloyes}},
  \bibinfo {author} {\bibfnamefont {S.}~\bibnamefont {Balle}}, \ and\ \bibinfo
  {author} {\bibfnamefont {M.}~\bibnamefont {Giudici}},\ }\href {\doibase
  10.1103/PhysRevLett.112.223901} {\bibfield  {journal} {\bibinfo  {journal}
  {Phys. Rev. Lett.}\ }\textbf {\bibinfo {volume} {112}},\ \bibinfo {pages}
  {223901} (\bibinfo {year} {2014})}\BibitemShut {NoStop}%
\bibitem [{\citenamefont {Camelin}\ \emph {et~al.}(2016)\citenamefont
  {Camelin}, \citenamefont {Javaloyes}, \citenamefont {Marconi},\ and\
  \citenamefont {Giudici}}]{CJM-PRA-16}%
  \BibitemOpen
  \bibfield  {author} {\bibinfo {author} {\bibfnamefont {P.}~\bibnamefont
  {Camelin}}, \bibinfo {author} {\bibfnamefont {J.}~\bibnamefont {Javaloyes}},
  \bibinfo {author} {\bibfnamefont {M.}~\bibnamefont {Marconi}}, \ and\
  \bibinfo {author} {\bibfnamefont {M.}~\bibnamefont {Giudici}},\ }\href
  {\doibase 10.1103/PhysRevA.94.063854} {\bibfield  {journal} {\bibinfo
  {journal} {Phys. Rev. A}\ }\textbf {\bibinfo {volume} {94}},\ \bibinfo
  {pages} {063854} (\bibinfo {year} {2016})}\BibitemShut {NoStop}%
\bibitem [{\citenamefont {Schelte}\ \emph {et~al.}(2018)\citenamefont
  {Schelte}, \citenamefont {Javaloyes},\ and\ \citenamefont
  {Gurevich}}]{SJG-PRA-18}%
  \BibitemOpen
  \bibfield  {author} {\bibinfo {author} {\bibfnamefont {C.}~\bibnamefont
  {Schelte}}, \bibinfo {author} {\bibfnamefont {J.}~\bibnamefont {Javaloyes}},
  \ and\ \bibinfo {author} {\bibfnamefont {S.~V.}\ \bibnamefont {Gurevich}},\
  }\href {\doibase 10.1103/PhysRevA.97.053820} {\bibfield  {journal} {\bibinfo
  {journal} {Phys. Rev. A}\ }\textbf {\bibinfo {volume} {97}},\ \bibinfo
  {pages} {053820} (\bibinfo {year} {2018})}\BibitemShut {NoStop}%
\bibitem [{\citenamefont {Javaloyes}(2016)}]{J-PRL-16}%
  \BibitemOpen
  \bibfield  {author} {\bibinfo {author} {\bibfnamefont {J.}~\bibnamefont
  {Javaloyes}},\ }\href {\doibase 10.1103/PhysRevLett.116.043901} {\bibfield
  {journal} {\bibinfo  {journal} {Phys. Rev. Lett.}\ }\textbf {\bibinfo
  {volume} {116}},\ \bibinfo {pages} {043901} (\bibinfo {year}
  {2016})}\BibitemShut {NoStop}%
\bibitem [{\citenamefont {Gurevich}\ and\ \citenamefont
  {Javaloyes}(2017)}]{GJ-PRA-17}%
  \BibitemOpen
  \bibfield  {author} {\bibinfo {author} {\bibfnamefont {S.~V.}\ \bibnamefont
  {Gurevich}}\ and\ \bibinfo {author} {\bibfnamefont {J.}~\bibnamefont
  {Javaloyes}},\ }\href {\doibase 10.1103/PhysRevA.96.023821} {\bibfield
  {journal} {\bibinfo  {journal} {Phys. Rev. A}\ }\textbf {\bibinfo {volume}
  {96}},\ \bibinfo {pages} {023821} (\bibinfo {year} {2017})}\BibitemShut
  {NoStop}%
\bibitem [{\citenamefont {Pimenov}\ \emph {et~al.}(2018)\citenamefont
  {Pimenov}, \citenamefont {Javaloyes}, \citenamefont {Gurevich},\ and\
  \citenamefont {Vladimirov}}]{PJG-PTRA-18}%
  \BibitemOpen
  \bibfield  {author} {\bibinfo {author} {\bibfnamefont {A.}~\bibnamefont
  {Pimenov}}, \bibinfo {author} {\bibfnamefont {J.}~\bibnamefont {Javaloyes}},
  \bibinfo {author} {\bibfnamefont {S.~V.}\ \bibnamefont {Gurevich}}, \ and\
  \bibinfo {author} {\bibfnamefont {A.~G.}\ \bibnamefont {Vladimirov}},\ }\href
  {\doibase 10.1098/rsta.2017.0372} {\bibfield  {journal} {\bibinfo  {journal}
  {Philosophical Transactions of the Royal Society of London A: Mathematical,
  Physical and Engineering Sciences}\ }\textbf {\bibinfo {volume} {376}}
  (\bibinfo {year} {2018}),\ 10.1098/rsta.2017.0372},\ \Eprint
  {http://arxiv.org/abs/http://rsta.royalsocietypublishing.org/content/376/2124/20170372.full.pdf}
  {http://rsta.royalsocietypublishing.org/content/376/2124/20170372.full.pdf}
  \BibitemShut {NoStop}%
\bibitem [{\citenamefont {New}(1974)}]{N-JQE-74}%
  \BibitemOpen
  \bibfield  {author} {\bibinfo {author} {\bibfnamefont {G.}~\bibnamefont
  {New}},\ }\href {\doibase 10.1109/JQE.1974.1145781} {\bibfield  {journal}
  {\bibinfo  {journal} {Quantum Electronics, IEEE Journal of}\ }\textbf
  {\bibinfo {volume} {10}},\ \bibinfo {pages} {115 } (\bibinfo {year}
  {1974})}\BibitemShut {NoStop}%
\bibitem [{\citenamefont {Rosanov}\ and\ \citenamefont
  {Khodova}(1988)}]{RK-OS-88}%
  \BibitemOpen
  \bibfield  {author} {\bibinfo {author} {\bibfnamefont {N.~N.}\ \bibnamefont
  {Rosanov}}\ and\ \bibinfo {author} {\bibfnamefont {G.~V.}\ \bibnamefont
  {Khodova}},\ }\href@noop {} {\bibfield  {journal} {\bibinfo  {journal} {Opt.
  Spectrosc.}\ }\textbf {\bibinfo {volume} {65}},\ \bibinfo {pages} {449}
  (\bibinfo {year} {1988})}\BibitemShut {NoStop}%
\bibitem [{\citenamefont {Vladimirov}\ \emph {et~al.}(1999)\citenamefont
  {Vladimirov}, \citenamefont {Fedorov}, \citenamefont {Kaliteevskii},
  \citenamefont {Khodova},\ and\ \citenamefont {Rosanov}}]{VFK-JOB-99}%
  \BibitemOpen
  \bibfield  {author} {\bibinfo {author} {\bibfnamefont {A.~G.}\ \bibnamefont
  {Vladimirov}}, \bibinfo {author} {\bibfnamefont {S.~V.}\ \bibnamefont
  {Fedorov}}, \bibinfo {author} {\bibfnamefont {N.~A.}\ \bibnamefont
  {Kaliteevskii}}, \bibinfo {author} {\bibfnamefont {G.~V.}\ \bibnamefont
  {Khodova}}, \ and\ \bibinfo {author} {\bibfnamefont {N.~N.}\ \bibnamefont
  {Rosanov}},\ }\href {http://stacks.iop.org/1464-4266/1/i=1/a=019} {\bibfield
  {journal} {\bibinfo  {journal} {Journal of Optics B: Quantum and
  Semiclassical Optics}\ }\textbf {\bibinfo {volume} {1}},\ \bibinfo {pages}
  {101} (\bibinfo {year} {1999})}\BibitemShut {NoStop}%
\bibitem [{\citenamefont {Doedel}\ \emph
  {et~al.}(1991{\natexlab{a}})\citenamefont {Doedel}, \citenamefont {Keller},\
  and\ \citenamefont {Kernevez}}]{DKK-I-IJBC-91}%
  \BibitemOpen
  \bibfield  {author} {\bibinfo {author} {\bibfnamefont {E.}~\bibnamefont
  {Doedel}}, \bibinfo {author} {\bibfnamefont {H.~B.}\ \bibnamefont {Keller}},
  \ and\ \bibinfo {author} {\bibfnamefont {J.~P.}\ \bibnamefont {Kernevez}},\
  }\href {\doibase 10.1142/S0218127491000397} {\bibfield  {journal} {\bibinfo
  {journal} {International Journal of Bifurcation and Chaos}\ }\textbf
  {\bibinfo {volume} {01}},\ \bibinfo {pages} {493} (\bibinfo {year}
  {1991}{\natexlab{a}})},\ \Eprint
  {http://arxiv.org/abs/https://doi.org/10.1142/S0218127491000397}
  {https://doi.org/10.1142/S0218127491000397} \BibitemShut {NoStop}%
\bibitem [{\citenamefont {Doedel}\ \emph
  {et~al.}(1991{\natexlab{b}})\citenamefont {Doedel}, \citenamefont {Keller},\
  and\ \citenamefont {Kernevez}}]{DKK-II-IJBC-91}%
  \BibitemOpen
  \bibfield  {author} {\bibinfo {author} {\bibfnamefont {E.}~\bibnamefont
  {Doedel}}, \bibinfo {author} {\bibfnamefont {H.~B.}\ \bibnamefont {Keller}},
  \ and\ \bibinfo {author} {\bibfnamefont {J.~P.}\ \bibnamefont {Kernevez}},\
  }\href {\doibase 10.1142/S0218127491000555} {\bibfield  {journal} {\bibinfo
  {journal} {International Journal of Bifurcation and Chaos}\ }\textbf
  {\bibinfo {volume} {01}},\ \bibinfo {pages} {745} (\bibinfo {year}
  {1991}{\natexlab{b}})},\ \Eprint
  {http://arxiv.org/abs/https://doi.org/10.1142/S0218127491000555}
  {https://doi.org/10.1142/S0218127491000555} \BibitemShut {NoStop}%
\bibitem [{\citenamefont {Doedel}\ \emph {et~al.}(2007)\citenamefont {Doedel},
  \citenamefont {Fairgrieve}, \citenamefont {Sandstede}, \citenamefont
  {Champneys}, \citenamefont {Kuznetsov},\ and\ \citenamefont {Wang}}]{AUTO}%
  \BibitemOpen
  \bibfield  {author} {\bibinfo {author} {\bibfnamefont {E.~J.}\ \bibnamefont
  {Doedel}}, \bibinfo {author} {\bibfnamefont {T.~F.}\ \bibnamefont
  {Fairgrieve}}, \bibinfo {author} {\bibfnamefont {B.}~\bibnamefont
  {Sandstede}}, \bibinfo {author} {\bibfnamefont {A.~R.}\ \bibnamefont
  {Champneys}}, \bibinfo {author} {\bibfnamefont {Y.~A.}\ \bibnamefont
  {Kuznetsov}}, \ and\ \bibinfo {author} {\bibfnamefont {X.}~\bibnamefont
  {Wang}},\ }\href@noop {} {\emph {\bibinfo {title} {AUTO-07P: Continuation and
  bifurcation software for ordinary differential equations}}},\ \bibinfo {type}
  {Tech. Rep.}\ (\bibinfo {year} {2007})\BibitemShut {NoStop}%
\bibitem [{\citenamefont {Kutz}\ \emph {et~al.}(1998)\citenamefont {Kutz},
  \citenamefont {Collings}, \citenamefont {Bergman},\ and\ \citenamefont
  {Knox}}]{KCB-JQE-98}%
  \BibitemOpen
  \bibfield  {author} {\bibinfo {author} {\bibfnamefont {J.}~\bibnamefont
  {Kutz}}, \bibinfo {author} {\bibfnamefont {B.}~\bibnamefont {Collings}},
  \bibinfo {author} {\bibfnamefont {K.}~\bibnamefont {Bergman}}, \ and\
  \bibinfo {author} {\bibfnamefont {W.}~\bibnamefont {Knox}},\ }\href {\doibase
  10.1109/3.709592} {\bibfield  {journal} {\bibinfo  {journal} {Quantum
  Electronics, IEEE Journal of}\ }\textbf {\bibinfo {volume} {34}},\ \bibinfo
  {pages} {1749} (\bibinfo {year} {1998})}\BibitemShut {NoStop}%
\bibitem [{\citenamefont {Nizette}\ \emph {et~al.}(2006)\citenamefont
  {Nizette}, \citenamefont {Rachinskii}, \citenamefont {Vladimirov},\ and\
  \citenamefont {Wolfrum}}]{NRV-PD-06}%
  \BibitemOpen
  \bibfield  {author} {\bibinfo {author} {\bibfnamefont {M.}~\bibnamefont
  {Nizette}}, \bibinfo {author} {\bibfnamefont {D.}~\bibnamefont {Rachinskii}},
  \bibinfo {author} {\bibfnamefont {A.}~\bibnamefont {Vladimirov}}, \ and\
  \bibinfo {author} {\bibfnamefont {M.}~\bibnamefont {Wolfrum}},\ }\href
  {\doibase http://dx.doi.org/10.1016/j.physd.2006.04.013} {\bibfield
  {journal} {\bibinfo  {journal} {Physica D: Nonlinear Phenomena}\ }\textbf
  {\bibinfo {volume} {218}},\ \bibinfo {pages} {95 } (\bibinfo {year}
  {2006})}\BibitemShut {NoStop}%
\end{thebibliography}
\end{document}